\documentclass[nonacm, sigconf]{acmart}
\renewcommand\footnotetextcopyrightpermission[1]{} 
\usepackage{hyperref}
\usepackage{booktabs} 
\usepackage{epstopdf}
\usepackage{enumerate}
\usepackage{enumitem}
\newcommand{\eg}{e.g.,}

\usepackage{float}
\usepackage{CJKutf8}
\usepackage[utf8]{inputenc} 
\usepackage[T1]{fontenc}
\usepackage{listings}
\usepackage{color}
\usepackage{xcolor}

\renewcommand{\paragraph}[1]{\vspace{5pt}\noindent\textbf{#1}}

\usepackage{adjustbox}
\newcommand{\tabincell}[2]{\begin{tabular}{@{}#1@{}}#2\end{tabular}}
\usepackage{url}

\usepackage{xspace}
\usepackage{multirow}

\newcommand{\id}{device ID\xspace}
\newcommand{\ids}{device IDs\xspace}

\newcommand{\ps}{privilege separation\xspace}

\usepackage{tikz}
\usepackage{subfigure}
\usepackage{stfloats}

\setcopyright{none} 

\begin{document}

\title{Logic Bugs in IoT Platforms and Systems: A Review}



\author{Wei Zhou}
\affiliation{
  \institution{UCAS; PSU}
}
\author{Chen Cao*}
\affiliation{
  \institution{PSU}
}
\author{Dongdong Huo*}
\affiliation{
  \institution{CAS}
}
\author{Kai Cheng*}
\affiliation{
  \institution{CAS; PSU}
}
\author{Lan Zhang*}
\affiliation{
  \institution{PSU}
}
\author{Le Guan*}
\affiliation{
  \institution{UGA}
}
\author{Tao Liu*}
\affiliation{
  \institution{PSU}
}
\author{Yaowen Zheng*}
\affiliation{
  \institution{CAS}
}
\author{Yuqing Zhang}
\affiliation{
  \institution{UCAS}
}
\author{Limin Sun}
\affiliation{
  \institution{CAS}
}
\author{Yazhe Wang}
\affiliation{
  \institution{CAS}
}
\author{Peng Liu}
\affiliation{
  \institution{PSU}
}

\begin{abstract}
In recent years, IoT platforms and systems have been rapidly emerging.
Although IoT is a new technology, new does not mean simpler (than existing networked systems).  
Contrarily, the complexity (of IoT platforms and systems) is actually being increased in terms of the interactions between the physical world and cyberspace.
The increased complexity indeed results in new vulnerabilities. 
This paper seeks to provide a review of the recently discovered
logic bugs that are specific to IoT platforms and systems. 
In particular, 17 logic bugs and one weakness falling into seven categories of vulnerabilities 
are reviewed in this survey. 
\end{abstract}


\maketitle
\renewcommand{\thefootnote}{\fnsymbol{footnote}} 
\footnotetext[1]{These authors contributed equally to this work.}

\section{Introduction}

Leveraging devices connected to the Internet, IoT (Internet of Things) platforms
and systems have been significantly enhancing the interactions between
the physical world and the cyberspace (e.g., clouds). 
These interactions not only enable users and enterprises to 
gain better situation awareness of the physical world 
events they care about, but also enable optimized actuation 
and physical effect generation (e.g., changing the temperature in a room). 

In recent years, IoT platforms and systems have been rapidly emerging. 
Taking smart homes as one example, according to Statista research, more 
than 45 million smart home devices were installed in 2018,  and  
the  annual  growth  rate  of  home  automation is 22\%~\cite{SH2018}. 
Taking enterprise IoT as another example, 
as stated in a recent survey conducted by Microsoft~\cite{IoT-Signals}, 
``The enthusiasm for IoT adoption is global, and it also crosses industries. 
Among the enterprise IoT decision makers we surveyed, 
85\% say they have at least one IoT project in either 
the learning, proof of concept, purchase, or use phase, with 
many reporting they have one or more projects 
currently in `use'.'' 

Although IoT is a new technology, new does not mean simpler 
(than existing networked systems).  
Contrarily, the complexity (of IoT platforms and systems) is actually being 
  increased along the following dimension:  
The IoT technology introduces not only a significant number of nodes (e.g., IoT devices) to the global information grid, but also a significant amount of various cyber-physical relationships.  
The increased complexity provides the adversary with not only new vulnerabilities to explore, but also new opportunities for attackers to compromise an IoT platform in a previously unexpected manner. 

Motivated by the above observation, 
this paper seeks to provide a review of the recently discovered security vulnerabilities that are specific to IoT platforms and systems. 
In particular, the review will be dedicated to recently discovered logic bugs in IoT platforms and systems. 
By ``logic bugs'', we mean the vulnerabilities directly associated with the 
design logic of (certain part of) an IoT platform/system. 
It should be noticed that since this paper aims to provide a dedicated review 
of logic bugs, low-level security bugs (e.g., buffer overflow 
vulnerabilities in IoT firmware) are out of the scope of the paper. 

The remaining of the paper is as follows. 
In Section 2, we will present the system model of a typical 
real-world IoT platform. 
In Section 3, we present a classification 
   of the recently discovered logic bugs in IoT platforms and systems. 
In Sections 4-10, we provide a review of seven categories of logic bugs, respectively. 
In Section 11, we comment on the lessons learned through this literature review.  
In Section 12, we conclude the paper.

\section{Background}
\begin{figure}[t]
\includegraphics[width=\columnwidth]{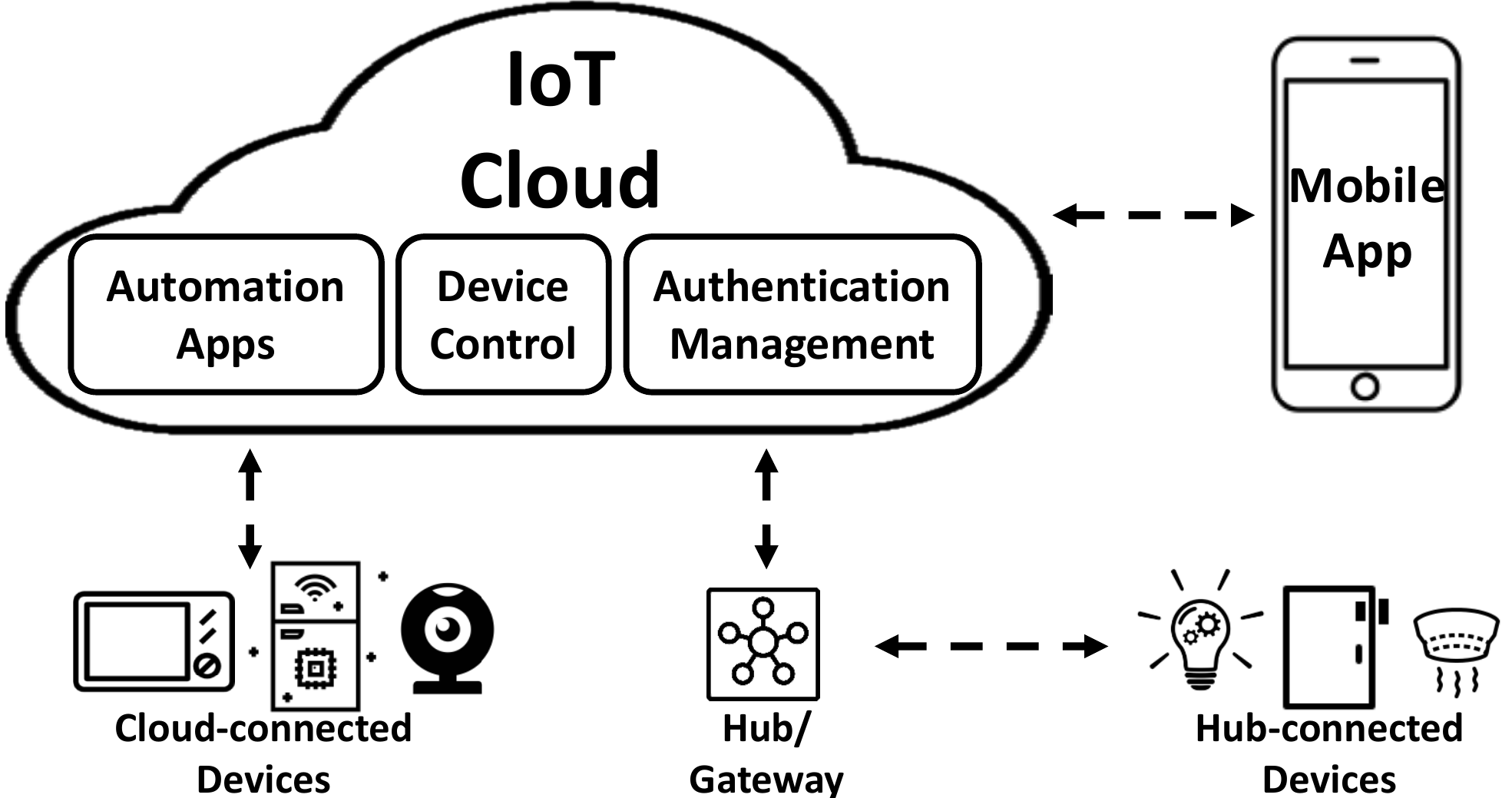}
\caption{Overview of IoT Platform Architecture} 
\label{fig:arch}
\end{figure}

\subsection{IoT Platform Architecture}
\label{sec:arch}
Although individual IoT platforms adopt
different strategies for commercialization, 
when it comes to the general design, all of them are very similar.
As shown in Figure~\ref{fig:arch},
typically, there are three major entities involved on an IoT platform:
IoT devices, the IoT cloud back-end, and the smartphone companion mobile app.

All the services provided by IoT devices is enabled by their brain — the IoT cloud.
Generally, it is usually responsible for three kinds of services, denoted as Authorization Management, Device Control, and Home Automation. 
First, in order to ensure that only the device owner and delegated users have access to their device,
the IoT cloud needs to 
authenticate the owner and device,
and maintain a one-to-one binding relationship between the owner's account and the device.
The binding relationship will be built when the device
is first set up.
Second, in order to allow users to control a device remotely,
the device control service serves as a ``proxy'' when users send remote commands to the device.
Lastly, most IoT platforms support home automation applications (i.e., IoT app), in which users can customize their own automation rules or install automation apps from official automation apps market (e.g., SmartThings's SmartApps) or third-party services (e.g., IFTTT).
These services are typically triggered by events.
The role of the IoT cloud is to check the permission of automation application
and to send control commands to the devices.

IoT devices equipped with embedded sensors and actuators collect physical states and events from the surrounding environment and send them directly or via a hub to the cloud.
According to how the devices interact with an IoT cloud, we can classify them as two types, namely cloud-connected devices and hub-connected devices.
WiFi-enabled devices can connect to the Internet
and thus most of them are cloud-connected devices that are designed to directly communicate with the IoT cloud.
Other energy-economic devices are not equipped with a WiFi interface. 
Instead, they need first connect to a hub/gateway using energy-efficient protocols such as Z-Wave and ZigBee. 
Then the hub connects to the IoT cloud on behalf of the IoT devices. 
We call the devices connected to a hub as hub-connected devices.
Note that although most IoT platforms support both kinds of devices,
IoT devices are designed to interact wit cloud only through the hub in some IoT platforms,
thus all these devices are hub-connected devices including WiFi-enabled devices.

The last kind of entity on an IoT platform is mobile apps.  
They provide users with an interface to setup device
(e.g., providing the WiFi credential), manage devices (e.g., binding a device with its owner's account) and install or create
home automation.

\subsection{IoT Device Bootstrapping}
\label{sec:lifecycle}
Before the device can be remotely controlled and
monitored by the authorized user, there are several steps to
setup the device.
Although IoT platforms use
different ways to implement the deployment of new devices, 
we found most of them are very similar.
Typically, this whole setup process is called IoT device bootstrapping.

\begin{enumerate}
\item \textbf{Device Discovery/Pairing}:
After the user has installed the official mobile app of the IoT platform,
he logs in the mobile app with his registered account and physically hard reset the device (e.g., pushing reset button).
Then he needs to discover the IoT device by scanning the QR code on the device label using the mobile app or manually selecting the target device model name listed in the mobile app.
The app will then broadcast the discovery message.
The target device responds by reporting to the app the basic device information such as MAC address,
device model, and firmware version.

\item \textbf{Internet Provisioning}: To access the Internet, WiFi-based devices can use several off-the-shelf mechanisms, including Access Point Mode~\cite{Chang2015Design}, WiFi Direct~\cite{WiFidirect} and SmartConfig~\cite{SmartConfig} to achieve WiFi provisioning.
Other types of devices using ZigBee or Bluetooth can indirectly connect to the Internet through hub or smartphone.

\item \textbf{Device Registration}:
The IoT cloud identifies a legitimate IoT device by a unique \id,
which is the most important identity information of a device.
The IoT platforms usually adopt the following two ways to provided \ids.
First, IoT devices send their unique information (\eg~MAC address, serial number)
and some legitimacy credential (\eg~embedded secret) to the cloud.
The cloud verifies the legitimacy and generates a \id, which is returned
to the device and written to the device's persistent storage.
The cloud also keeps the \id for future authentication.
Second,
some IoT platform providers who also fabricate their own device
usually generate
the \ids beforehand and hard-coded into the devices during fabrication.

\item \textbf{Device Binding}:
The IoT cloud binds the \id with the user account.
Note that binding request could be directly sent by mobile app or forwarded by device. 
As a result, only the authorized user can access the device via the cloud.
If other users request to bind the same device again,
the cloud will refuse this request unless the device has already been unbound.

\item \textbf{Device Login}:
The device uses the \id to log in to the cloud.
The cloud then marks the device as online and synchronizes the online state of the device to the mobile app.
The device and the cloud then maintain a heartbeat connection
to keep the device status periodically updated.

\item \textbf{Device in Use}:
In this phase, the device can interact with the cloud to perform the tasks.
In addition, the user can monitor the real-time status of the device and explicitly send control commands remotely via the mobile app.

\end{enumerate}

Note that the order of steps (1) and (2)
is exchangeable depending on the concrete implementation.

\begin{figure}[t]
\includegraphics[width=\columnwidth]{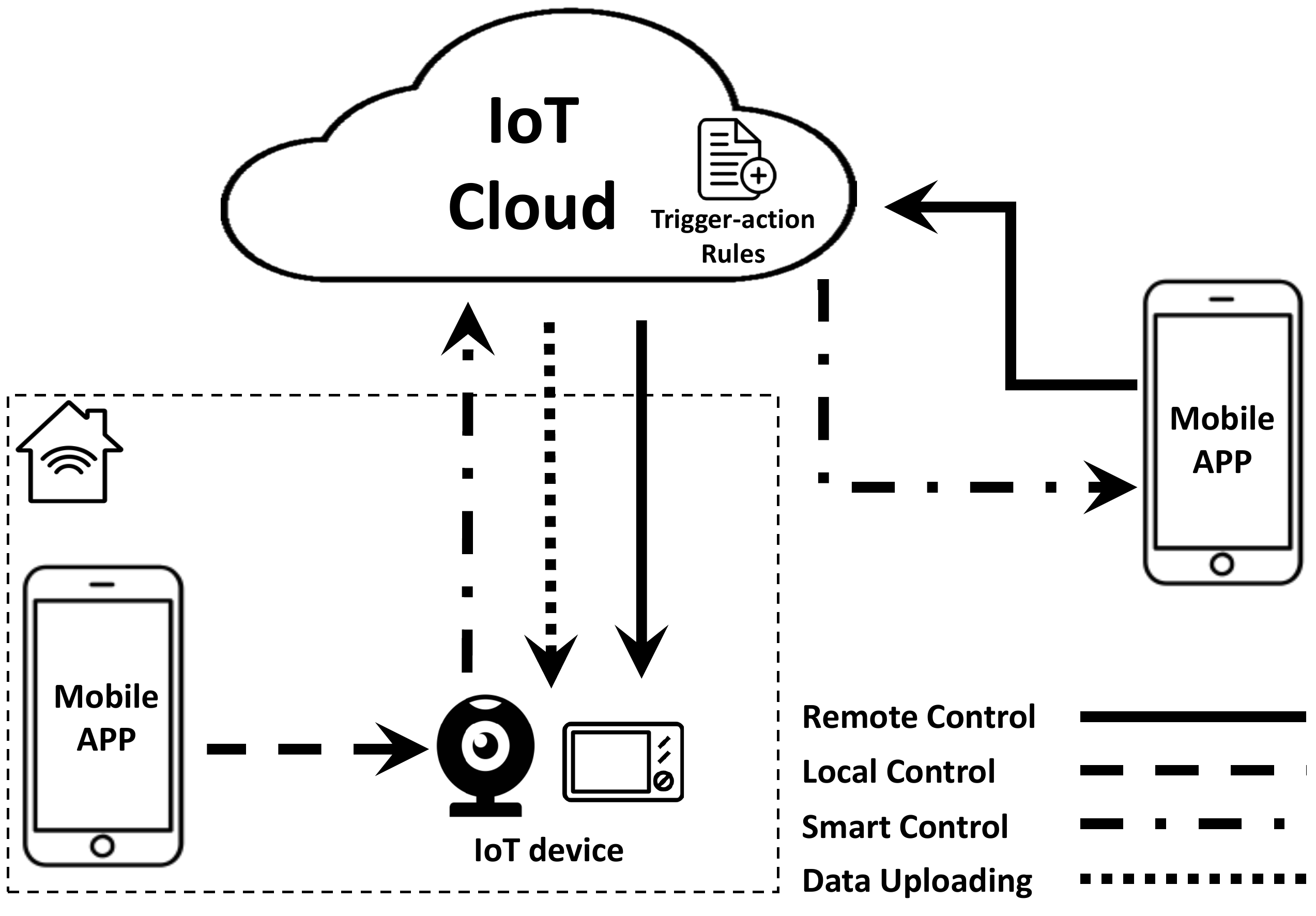}
\caption{IoT Device Communication Channels} 
\label{fig:commodel}
\end{figure}

\subsection{Communication Model}
\label{sec:comm}
Typically, 
IoT device interacts with cloud or mobile in four types of communication channels.
The channels serve for either device control or data transmission, and operate either
under crypto protection or in plaintext.
We depict them in Figure~\ref{fig:commodel}.

\begin{enumerate}
\item \textbf{Remote Control:}
When the user's smartphone can access the Internet,
he can remotely monitor and control his device via IoT cloud.
Specifically, the mobile app sends control commands of the target device identified by
\id to the IoT cloud.
Then the IoT cloud checks whether the user is authorized to access this device.
If the checking is passed, the cloud forwards the command to the target device.

\item \textbf{Local Control:}
When the user is in the same LAN with the device, he can directly send
local control commands to the target device.
Some IoT platforms such as Ali's Alink are exceptions because they
stick to the remote control channel even if the mobile app and the device are in the same LAN.

\item \textbf{Smart Control:}
The IoT cloud can automatically send control commands
to the device when automation rules are satisfied.
For example, the user can edit a rule that turns on the smart plug at a specified time
if the temperature is below 70$^\circ$F.
The rule is synchronized to the cloud.
When the time comes and thermometer indicates that the temperature is below 70$^\circ$F,
the cloud will automatically send a ``turn on'' command to the smart plug.


\item \textbf{Data Uploading:}
The IoT devices typically upload three kinds of messages (e.g., command response, heart-beat message and event notifications) to the IoT cloud.
First, the device need to reply to the control commands to notify cloud if the commands have been successfully executed and some commands also ask the device to continually upload their sensor data (e.g., video recording).
Second, the IoT devices routinely send heart-beat message to the cloud, 
thus the cloud can be aware of the connection with the device.
Note that many platforms also include current device status into the heart-beat message,
so that
the user can monitor the latest device status remotely.
Lastly, IoT devices also need to upload the event notifications (e.g., motion detection) to the IoT cloud
so that the cloud can trigger smart control commands to the smart home.


\end{enumerate}




\section{A Preliminary Classification}

\begin{table*}[t]
  \caption{A Preliminary Classification of IoT Logic Bugs}
  \begin{center}
  \begin{adjustbox}{width=\textwidth}
  \begin{tabular}{c|l|l}
    \toprule
    \textbf{Category}&\hspace{8 mm}\textbf{Ref}&\hspace{37 mm}\textbf{Name}\\
    \midrule
    \multirow{3}{*}{\tabincell{c}{\textbf{Authentication Problems}\\\textbf{Section~\ref{sec:bug1-3}}}}&Zhou19~\cite{zhou2019discovering}&\textbf{Bug 1:} Weak Device Authentication\\
    &Firmalice15~\cite{shoshitaishvili2015firmalice}&\textbf{Bug 2:} Device Authentication Bypass\\
    &Sethi19~\cite{misbinding}&\textbf{Bug 3:} Weak Owner Authentication\\
    \hline
    \multirow{3}{*}{\tabincell{c}{\textbf{Over-privileged Capability Management}\\\textbf{Section~\ref{sec:bug4-6}}}}&Fernandes16\cite{fernandes2016security}&\textbf{Bug 4:} Over-granted Capabilities in Automation Application\\
    &Fernandes~\cite{fernandes2016security}&\textbf{Bug 5:} Coarse-grained Capabilities in Automation Application\\
    &Yao19~\cite{yao2019identifying}&\textbf{Bug 6:} Privilege Separation Logic Bugs in IoT Firmware\\
    \hline
    \multirow{2}{*}{\tabincell{c}{\textbf{Working State Out of Synchronization}\\\textbf{Section~\ref{sec:bug7-8}}}}&Zhou19~\cite{zhou2019discovering}&\textbf{Bug 7:} Insufficient State Guard\\
    &Zhou19~\cite{zhou2019discovering}&\textbf{Bug 8:} Illegal States Combination\\
    \hline
    \multirow{2}{*}{\tabincell{c}{\textbf{Sensor Data Out of Synchronization}\\\textbf{Section~\ref{sec:bug9-10}}}}&Ocon19~\cite{oconnor2019blinded}&\textbf{Bug 9:} Sensor Blinding\\
    &Ocon19~\cite{oconnor2019blinded}&\textbf{Bug 10:} State Confusion\\ 
    \hline
    \multirow{4}{*}{\tabincell{c}{\textbf{Unexpected Trigger Action in Automation APP}\\\textbf{Section~\ref{sec:bug11-14}}}}&Celik18~\cite{celik2018soteria}&\textbf{Bug 11:} Race Conditions of Events\\
    &Celik18~\cite{celik2018soteria}&\textbf{Bug 12:} Attributes of Conflicting Values\\
    &Celik18~\cite{celik2018soteria}&\textbf{Bug 13:} Attributes Duplication\\
    &Celik18\cite{celik2018soteria}&\textbf{Bug 14:} Missing Events\\
    \hline
    \multirow{2}{*}{\tabincell{c}{\textbf{Information Flow Hijacking in Automation APP}\\\textbf{Section~\ref{sec:bug15-16}}}}&Bastys18~\cite{bastys2018if}&\textbf{Bug 15:} URL-based JS Injection\\&Bastys18~\cite{bastys2018if}&\textbf{Bug 16:} URL-based HTML Tag Injection\\
    \hline
    \multirow{2}{*}{\tabincell{c}{\textbf{Vulnerable Task Management in RTOS}\\\textbf{Section~\ref{sec:bug17}}}}&Dong19~\cite{openreview:LIPS}&\textbf{Bug 17:} Lack of Isolation between Context Table and Tasks\\
    &Dong19~\cite{openreview:LIPS}&\textbf{Weakness:} Inadequate Task Memory Isolation\\
    \bottomrule
  \end{tabular}
  \end{adjustbox}
  \end{center}
\label{tab:class}
\end{table*}

We classify the collected logic bugs
into seven categories based on their root causes.
In Table~\ref{tab:class}, we list the categories,
the corresponding logic bugs, a brief description for each bug, and
the corresponding references.
Note that this classification is preliminary,
because with the development of IoT technologies,
we expect the emergence of previously-unseen new vulnerabilities.

\paragraph{Authentication Problems.}
Authentication is a classic issue in systems security.
IoT systems are no exception~\cite{shoshitaishvili2015firmalice,zhou2019discovering,misbinding}.
More specifically, there are logic bugs in which the IoT devices
are mistakenly recognized as other devices (bug 1 and bug 3).


\paragraph{Over-privileged Capabilities.}
Similar to Android applications, IoT platforms also use capabilities to define and manage the privileges of  
the automation apps in the cloud.
However, recent research~\cite{fernandes2016security} disclosed that the automation apps are often granted more privileges than necessary and
the privileges were abused by attackers (bug 4 and bug 5).
Meanwhile, IoT devices also lack necessary privilege separation (bug 6).

\paragraph{Working State Out of Synchronization.}
The IoT devices, mobile apps and the IoT cloud interact with each other
closely in IoT platforms.
A critical event will cause a working state change in either of the
three entities.
Formally, the working state changes can be modelled as
a state machine.
However, the state machine transitions are often not
properly safeguarded in popular IoT platforms.
When an unexpected transition is triggered by attackers,
serious consequence could happen (bug 7 and bug 8).

\paragraph{Sensor Data Out of Synchronization.}
Due to intermittent network conditions,
the delivery of sensory measurements from the IoT devices
to the IoT cloud could be interrupted or delayed.
That will make the sensory data out of synchronization between IoT devices and the cloud.
Research~\cite{oconnor2019blinded} demonstrates how attackers can utilize this vulnerability to 
cause security hazards (bug 9 and bug 10).

\paragraph{Unexpected Trigger Action.}
The trigger action model is widely used in IoT automation apps.
Researchers~\cite{oconnor2019blinded} discover several logic bugs (bug 11-14) caused by unexpected trigger action chains.

\paragraph{Information Flow Hijacking in Automation APP.}
IoT platforms allow users to install third-party trigger-action services like IFTTT.
However,
as revealed in bug 15 and bug 16,
attackers can stealthily inject JavaScript code or HTML tags in malicious automation apps.
This redirects the action to the server which is 
under the control of attackers.

\paragraph{Vulnerable Task Management in RTOS.}
RTOSs are widely used in the resource-constrained IoT devices. 
Some RTOSs (e.g.,~Arm Mbed OS) feature task isolation for increased system security.
However, researchers found that there has a serious design flaw (e.g., bug 16) that can be exploited to bypass this protection~\cite{openreview:LIPS}.
In addition, constrained by the capability of the hardware,
no page-based memory protection can be supported in RTOSs.
Instead, the RTOS kernel and tasks often share the same
flat address space.
As a result, a simple memory error such as a buffer overflow in a vulnerable
task could corrupt the memory of another task or even the kernel.
Since this problem is not directly related to the IoT platform design logic,
we consider it as a weakness.

In the following, we detail these seven categories.
In each section,
we elaborate the logic bugs from five aspects.

\vspace*{-1mm}
\begin{enumerate}[label=\textbf{\Alph*}]
\item \textbf{System Model} describes
the technical background behind this logic bug;

\item \textbf{Attack Scenario}
describes the prerequisites of the attack,
how the attackers exploit this logic bug,
and the consequence of the attack;

\item \textbf{Cause Analysis} discusses 
the fundamental cause of why this vulnerability exists;


\item \textbf{Identifying Method}
describes the method used in identifying this logic bug;

\item \textbf{Defense} discusses 
how to defend against and mitigate this logic bug in the first place. 

\end{enumerate}





\section{Authentication Problems}
\label{sec:bug1-3}
\subsection{Bug 1: Weak Device Authentication}
\label{sec:bug1}
\paragraph{A. System model.}
To manage devices and provide remote service for users,
the IoT cloud needs to perform authentication checks on both users and devices.
Comparing to developed mobile-side user authentication,
the manufacturers and IoT platform providers deploy simple or no authentication for IoT devices.
Typically,
the device-cloud communication adopts one-way SSL protocol
and only the server certificates are hard-coded in the firmware.
That means the device only authenticates the cloud/server certificate,
but the cloud cannot authenticate the device via client certificate.

Thus, to realize device authentication,
some manufacturers hard-coded the server credential, the MAC address, serial number, device ID, etc. in the firmware.
Before building a connection with the device,
the cloud will check whether the information is legitimate or not.
Other companies use their own proprietary protocol for device-cloud communication.
They usually use hard-coded communication key or secrets used generated communication key in the firmware for device authentication.

\paragraph{B. Attack Scenario.}
For IoT platforms using communication key or secrets used to generate communication key for device authentication,
once these keys or secrets has leaked,
attackers can decrypt the device communication traffic and carry out man-in-the-middle (MITM) attack to the devices.
For IoT platforms using additional device information including device credentials for device authentication,
even if such information has leaked,
the attackers are still unable to decrypt the communication. 
However, Zhou et al.~\cite{zhou2019discovering} show that
attackers can leverage it to emulate non-existing devices to log in and keep the connection with the cloud.
Cloud will consider it as a real device.
The attackers can take advantage of that with other logic bugs as we shall see in Section~\ref{sec:bug7-8} to
intervene in the normal interactions of real devices.

\paragraph{C. Cause Analysis.}
The information used for device authentication
should be well-protected,
but actually
it is readily acquired by attackers in the real world.
First, some information is publicly available or can be
easily inferred. 
For example,
the attackers can guess
or brute-force attack device MAC address,
because the first three bytes in a MAC
address are usually fixed for a manufacturer.
Thus, the adversary can fix these
bytes and mutate the last 3 bytes.
In addition,
some IoT platform providers like Ali allow one credential to be used by multiple device authentication.
Even worse,
Zhou et al.~\cite{zhou2019discovering} also found there are a bunch of credentials used for Ali's IoT device authentication that are available on the official Github repositories of both the Ali company
and the cooperative manufacturers.

Some other information like communication keys makes
a brute-force guessing to them impossible.
However, such information is usually hard-coded and cannot
be changed once it is programmed.
Thus, once the attackers have physical access to 
the victim device,
such information become leaked forever.
Furthermore, compared to PC and mobile phone,
there are more
circumstances in which a victim use a
device which was once possessed by an attacker.
First, the consumer ownership of a device can be
changed if the device gets resold or decommissioned.
Second, the smart home device can be
shared with others in many scenarios such as vacation rentals and
hospitality services like Airbnb.
In both cases, the attackers
have a chance to extract device authentication information from the
device. 

\paragraph{D. Identifying Method \& Defense.}
The IoT platform should deploy strict device authentication mechanisms. 
Depending on the computation capability of the device,
for high-end IoT devices powered by high-end CPU like ARM Cortex-A,
the unique client certificate should be encrypted and stored into the One Time Programmable (OTP) register rather than firmware.
The cloud should adopt two-way SSL authentication and always check the client certificate. 
For resource-restricted IoT devices, 
the manufacturer should embed a read-only random number into the device. 
The cloud should always check whether the random number matches other unique device information like MAC address.

\subsection{Bug 2: Device Authentication Bypass}
\label{sec:bug2}
\paragraph{A. System model.}
In this system model, 
we consider the authentication bypass vulnerability only in the IoT device itself,
and it has nothing to do with the IoT Platform as mentioned in Section~\ref{sec:arch}.
These IoT devices provide a mini web server or a customized server with a listening port that
allows users to access and control the device through a web browser or directly through the listening port.
Certainly,
some sensitive operations of IoT devices can only be performed by authorized users. 
Taking a network camera as an example,
only authorized users have permission to watch the recorded video and change the recording settings.
Thus, IoT devices protect these privileged operations through user verification.
The typical verification mechanism is to check the username and password stored in IoT devices.
Before the user can operate the device, a pair of username and password will be required.
Then, the device performs authorization verification by comparing the credentials stored in the device with the password provided by the user.
Such an authentication process is implemented in the firmware of the device.

\paragraph{B. Attack Scenario.}
Authentication bypass vulnerability, commonly termed ``backdoors'', allows an attacker to execute 
privileged functionalities (e.g., password modification, video downloads, and firmware upgrades, etc.) without knowing the valid credentials of an authorized user.
For example, Santamarta presented a backdoor attack to the Schneider ION 8600 smart meter at BlackHat in 2012~\cite{backdoorOne}.
He found a Factory Login account as ``reserved'' by reading the meter's documents.
Then, he reverse-engineers the firmware of the smart meter and discovered a factory login account that allows an attacker to fully control the device.
This account is a 32-bit number that could be computed using a hash algorithm seeded with a hard-coded ``secret'' string and 
the serial number of the smart meter.
Therefore, an attacker can access the smart meter via telnet to obtain the serial number and generate the factory login account.
Then, the attacker can use this account to modify protected data such as billing.

\paragraph{C. Cause Analysis.}
There are three reasons for the authentication bypass bug. 
The first reason is the intentionally hard-coded credentials.
Some manufacturers hard-coded credentials that are unknown to the user for device maintenance and upgrade.
For example, the backdoor in the smart meter is a hard-coded credential.
The second reason is the intentionally hidden authentication interface.
Such interfaces do not require authorization to access the privileged operations in IoT devices.
The third reason is that the unintended bugs compromise the integrity of the authentication routine or bypass it entirely.

\paragraph{D. Identifying Method \& Defense.}
Shoshitaishvili et al. have presented Firmalice~\cite{shoshitaishvili2015firmalice}, a binary firmware analysis framework to discover the authentication bypass bug.
First, Firmalice converts the firmware binary to an intermediate language called VEX, discovers the entry point and identifies privileged program points.
Then, Firmalice uses code slicing techniques to extract code snippets associated with the privileged program point, relieving symbolic execution path explosion problems.
Finally, Firmalice performs symbolic execution on the sliced code and 
attempts to solve the path constraints at the privileged point to concretize the user input.
If the user input can be uniquely concretized, then it represents that the input required to reach the privileged program point can be uniquely determined by the attacker, and the associated path is labeled as an authentication bypass.
Since most authentication bypass bugs in IoT devices are backdoors that deliberately left by firmware developers, 
we think the best defense is to patch and upgrade the firmware
by discovering authentication bypass early through program analysis techniques (e.g., Firmalice).






\subsection{Bug 3: Weak Owner Authentication}
\label{sec:bug3}

\paragraph{A. System model.}
Some IoT device manufacturers do not deploy their devices with IoT platform,
so that they have to adopt other protocols like
Nimble out-of-band authentication for Extensible Authentication Protocol (EAP-NOOB) to implement
bootstrapping of new devices.

\begin{figure}[h]
\includegraphics[width=\columnwidth]{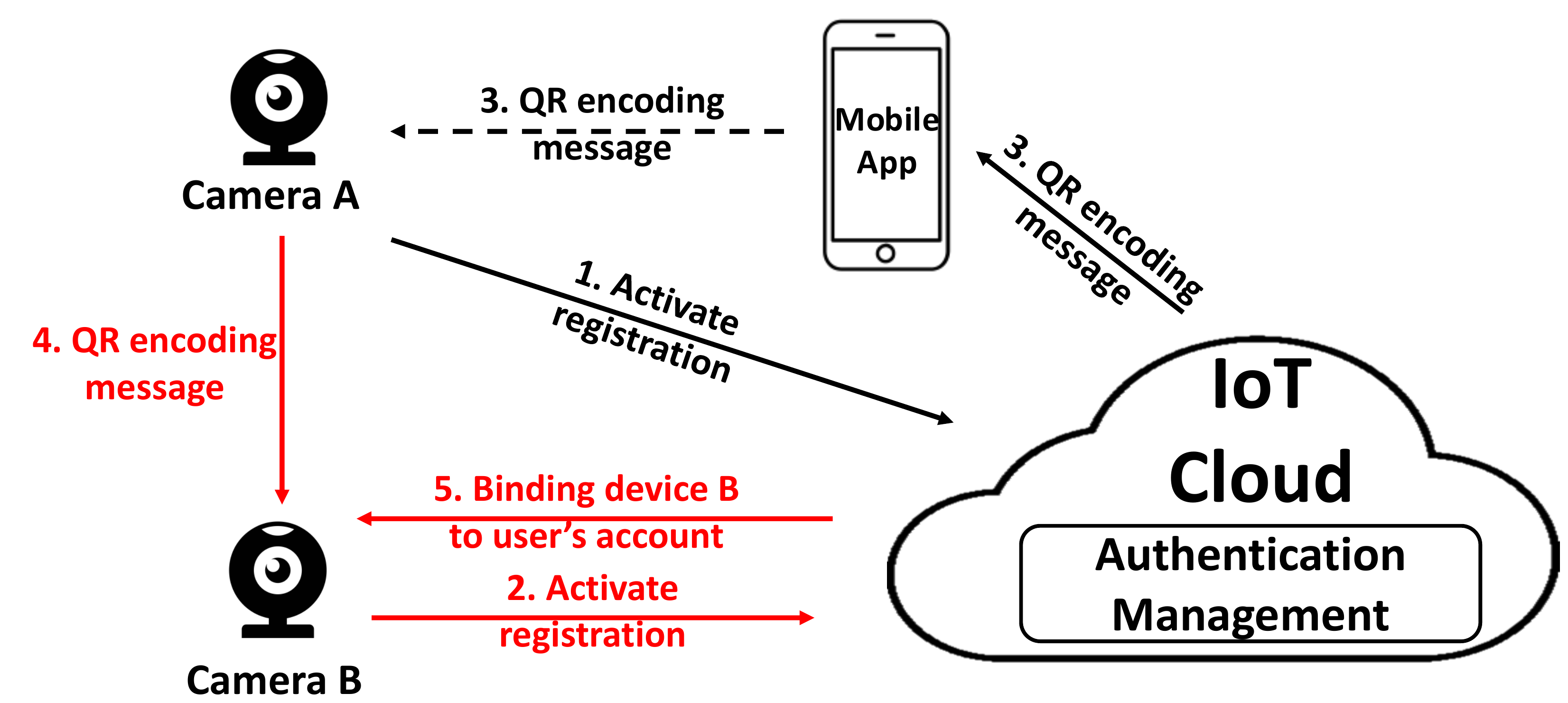}
\caption{Attack Scenario of Bug 3}
\label{fig:misbinding}
\end{figure}

For EAP-NOOB protocol,
a human-assisted-out-of-bind (OOB) channel is added 
to achieve device binding process.
Specifically, when a user wants to bind the device, he first needs to deliver his user authentication message to the device in an OOB channel.
The form of user authentication message could be QR code, audio signal, NFC data, etc.
Then the device transmits user authentication message to the cloud with its identity information,
finally the cloud can bind this device with the user's account.
For example, when the user binds the camera, the IoT cloud will generate a QR code associated with his account and send it to user's mobile app.
Then user should let the camera scan the QR code to complete the device binding process. 
\paragraph{B. Attack Scenario.}
We show a specific attack scenario in Figure~\ref{fig:misbinding}.
The user first resets the device to activate the device registration.
At the same time,
the attacker also activates the registration of device B.
After that, the user logs in his account and choose the camera A, and the QR code encoding as authentication messages are generated from the cloud. 
Then, the user shows the QR code to the camera A (in the OOB channel indicated by the dashed line in Figure~\ref{fig:misbinding}).
Since the device A is compromised and controlled, 
the attacker would deliver the message received by the device A to another device B.
With the authentication message, the attacker successfully binds the camera B to the user's account.
As a result, the device B owns authentication message and would be successfully associated with the user's account on the IoT cloud which is against the user's intention.
Researchers~\cite{misbinding} call this attack as misbinding attack.

\paragraph{C. Cause Analysis.}
In the IoT device bootstrapping process, the IoT cloud lacks adequate authentication of target IoT device which causes this logic bug.
Specifically, the IoT cloud associates the user's account with the IoT device which provides the user's authentication message generated by the mobile app.
Once the target IoT device is compromised, 
the information could be stolen by the attacker and used for another device binding.

In addition, IoT devices take user's physical access to devices as their identities instead of cryptographically verifiable identities such as serial number, public keys, which makes it hard for the IoT cloud to authenticate the IoT device.


\paragraph{D. Identifying Method.}
To analyze an authentication protocol whether this logic bug exists or not in it, 
Sethi et al.~\cite{misbinding} have proposed a formal model analysis approach based on an automatic cryptographic protocol verifier named Proverif.
This analysis approach finally finds two forms of misbinding. 
One is shown in the attack scenario, and another is that both devices are compromised and the bug could be exploited similarly.


\paragraph{E. Defense.}
Approaches such as identifier communication and presence checking have been proposed by Sethi et al.~\cite{misbinding} to partially defend the attacks.
In the identifier communication approach, 
the IoT cloud utilizes some printable information such as model, serial numbers and even public-key fingerprint attached to the device for enhancing device authentication. 
Thus, it would be more difficult for attackers to launch the misbinding attack.
In the presence checking approach,
the user always communicates with the dynamic root of trust for measurement (DRTM) inside the device and generates authentication approaches based on trust computing base (TCB), which could check the presence of the device correctly even with untrusted software in the IoT device.






\newcommand{\sapp}{automation application\xspace}


\section{Over-privileged capabilities management}
\label{sec:bug4-6}

\subsection{System Model}
Recently, many IoT platform providers open their automation application programming frameworks to support third party IoT apps development.
These programming frameworks usually define a set of capabilities to manage the permissions of IoT apps.
A capability in the IoT platform consists of a set of commands (i.e., method calls) and attributes (i.e., properties)~\cite{fernandes2016security}.
Commands represent ways in which a device can be controlled.
Attributes represent the state information of a device. 
When installing an automation application in the IoT platform, the user would be asked to authorize what capabilities to this application.
Once it is installed, the application can send commands and obtain attributes to/from the related devices, bound with these capabilities.

\subsection{Bug 4: Over-granted Capabilities in Automation Apps}
\label{sec:bug4}
\paragraph{A. Attack Scenario.}
A malicious \sapp advertises itself as a battery status monitor application.
However, it requests a set of capabilities which is beyond the requirements of a battery status monitor application, including \emph{capability.lock} and \emph{capability.unlock}.
When being installed into the system, this application is authorized the capabilities of monitoring the battery status of the front door lock as well as locking and unlocking it. 
As a result, this \sapp can secretly unlock the door without the user's attention.
Hence, it puts the user under the threat of break-ins and theft.

\paragraph{B. Cause Analysis.}
An \sapp can request capabilities beyond what its advertisement describes.
However, from the description, the user usually has no knowledge of what capabilities can be abused when he chooses to install the \sapp.
The root cause of this bug is the gap between what capabilities would be used in the user's mind and the reality of what capabilities can be used.

\paragraph{C. Identifying Method \& Defense.}
Tian et al. proposed a tool, SmartAuth, identifying the \sapp which requests more capabilities and has more functionalities than its advertising~\cite{tian2017smartauth}.
It collects security-relevant information from the \sapp's description, code and annotations, and identifies discrepancies between what is claimed in the description and what the app actually does.
Then the information is used to inform the user how the specific application has the inconsistency between its description and its code.
Therefore, the users have the knowledge to determine whether the \sapp can abuse certain capabilities and enforce whether the \sapp can utilize certain capabilities with SmartAuth through different security policies.
Because most IoT platforms are closed-source, SmartAuth patches the \sapp to implement the proof-of-concept system.
In the end, each \sapp can only access what the user allows.

\subsection{Bug 5: Coarse-grained Capabilities in Automation Apps}
\label{sec:bug5}

\paragraph{A. Attack Scenario.}
When being installed into the IoT platform, a benign-but-buggy or malicious \sapp requests to use only one command \emph{lock} of \emph{capability.lock}.
Because of the coarse-grained capabilities in the platform, this capability also includes command \emph{unlock}.
That means the \sapp has the ability to automatically send \emph{unlock} command.
If this capability is bound to a front door lock, this \sapp can lock and unlock the front door.
Hence, when this benign-but-buggy or malicious \sapp is exploited by an attacker, she can unlock the front door, which can result in break-ins or theft.

\paragraph{B. Cause Analysis.}
The root cause of this bug is the coarse-grained capabilities classification in the automation application programming frameworks.
One capability may include one or more commands and attributes.
Once an \sapp requests one command, a set of other commands or attributes included in one capability are also authorized to this app automatically.

\paragraph{C. Identifying Method.}
The method of identifying whether an \sapp has been authorized more commands or attributes than it requires is to verify the result of {\textit{requested commands and attributes} - \textit{used commands and attributes}}.
If it is empty, \sapp is not over-privileged.
The \textit{requested commands and attributes} can be directly obtained from the capabilities requested by the \sapp.
To get the \textit{used commands and attributes}, Fernandes et al. utilize static analysis to determine a list of all methods and properties accessed in an \sapp~\cite{fernandes2016security}.
Then this list is filtered using the completed capability documentation to obtain the set of used commands and attributes in this app.
In the end, the set of over-authorized commands or attributes can be computed.

\paragraph{D. Defense.}
The root cause is essentially the design flaw of the IoT platform.
Therefore, to defend the attacks caused by the coarse-grained capabilities,
the design of this system should be re-constructed.
However, most IoT platforms are closed-source.
Moreover, the cost of using a new design may be huge because of deployed devices and applications.
Taking this into consideration, patching the \sapp could be the only solution to defend this kind of attack.
Jia et al. propose ContextIoT, which can automatically patch unmodified commodity \sapp, to provide fine-grained capabilities in the IoT platform~\cite{jia2017contexlot}.
ContextIoT consists of two major steps at two stages, e.g., installation time and runtime.
At the installation time, ContextIoT collects context information from the \sapp and patches it to separate security sensitive behaviors (e.g., unlock) which request permissions from the user, if the context is not logical.
At runtime, ContextIoT prompts the request to the user to ask for permission if the behavior does not conform with a certain security logic.
Essentially, ContextIoT prevents the usage of capabilities authorized to an \sapp for malicious behaviors.

\subsection{Bug 6: Privilege Separation Logic Bugs in IoT Firmware}
\begin{figure}[t]
\includegraphics[width=\columnwidth]{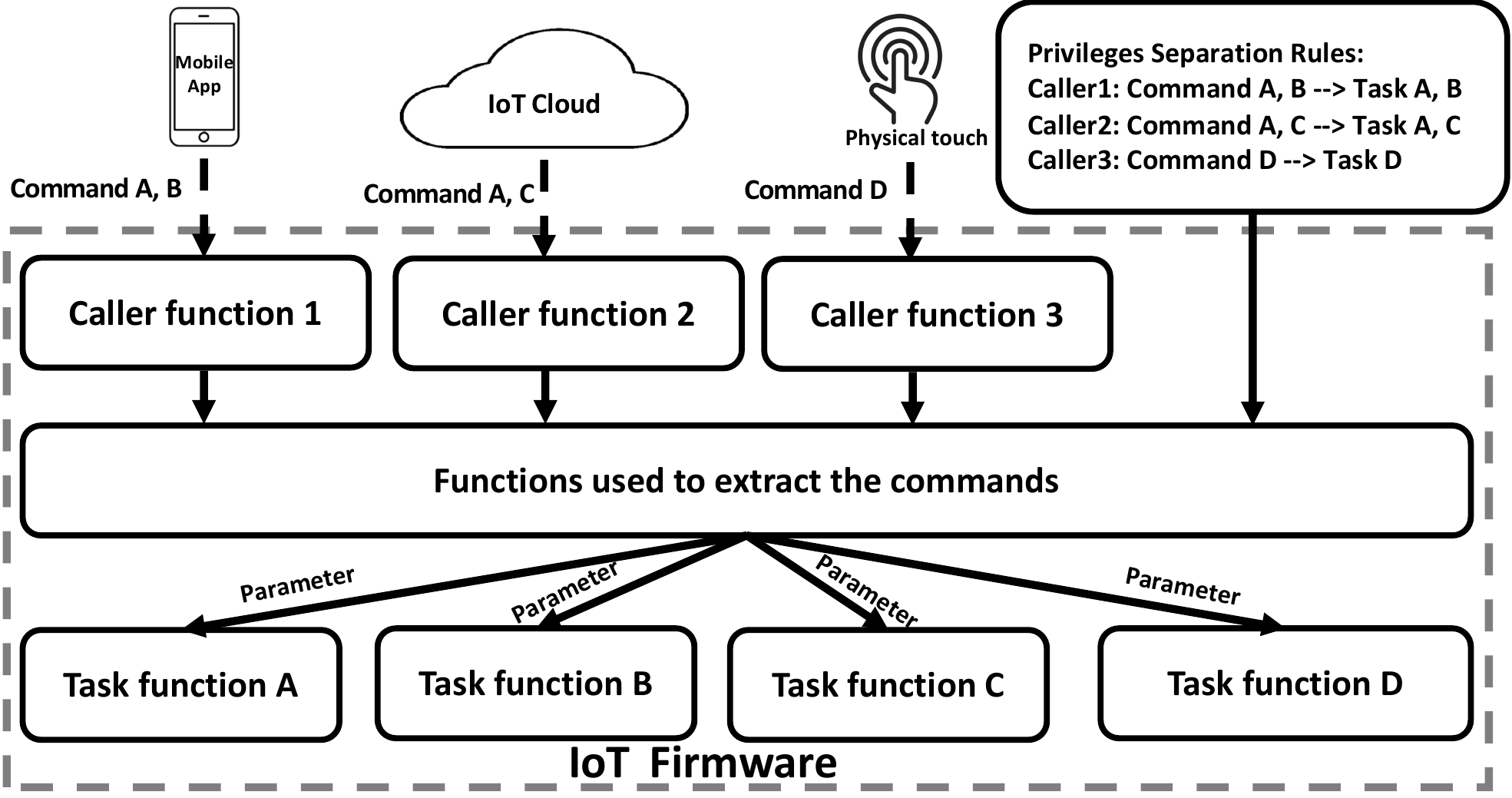}
\caption{System Model of Bug 5} 
\label{fig:bug12m}
\end{figure}

\label{sec:bug6}
\paragraph{A. System Model.}
IoT devices continuously interact with different entities including mobile \sapp, cloud or physical access (e.g., pushing a button) and
perform the tasks corresponding to
the user commands.
As shown in Figure~\ref{fig:bug12m},
because different communication channels usually use different protocols, ports and servers,
the IoT firmware images are implemented with different functions to receive and decode the message from
different interactive entities.
We name the first function used to receive the message from interactive entities as caller functions.
After decoding the message,
IoT firmware images invoke other functions to extract the commands and finally trigger the corresponding functions to accomplish the specific tasks.
We name the first function used to perform tasks for individual command as task function.

In addition, since different entities play distinct roles in an IoT platform,
the command sets from these diverse interactive entities are differential. 
That means 
some commands could only be invoked by specific 
interactive entities in normal operations.
For instance,
remote commands sent by the cloud are usually
responsible for device management services like assigning device identification.
Thus,
most functions in IoT firmware can be divided into separated collections for dealing with commands invoked by different entities.
Each collection should have distinct privilege,
so that one entity can only invoke its own commands.
For example,
according to the privilege separation rules in Figure~\ref{fig:bug12m},
Task function B should only be invoked by command B sent by mobile app and
Task function C should only be invoked by command C sent by cloud and so on.

\begin{figure}[t]
\includegraphics[width=0.9\columnwidth]{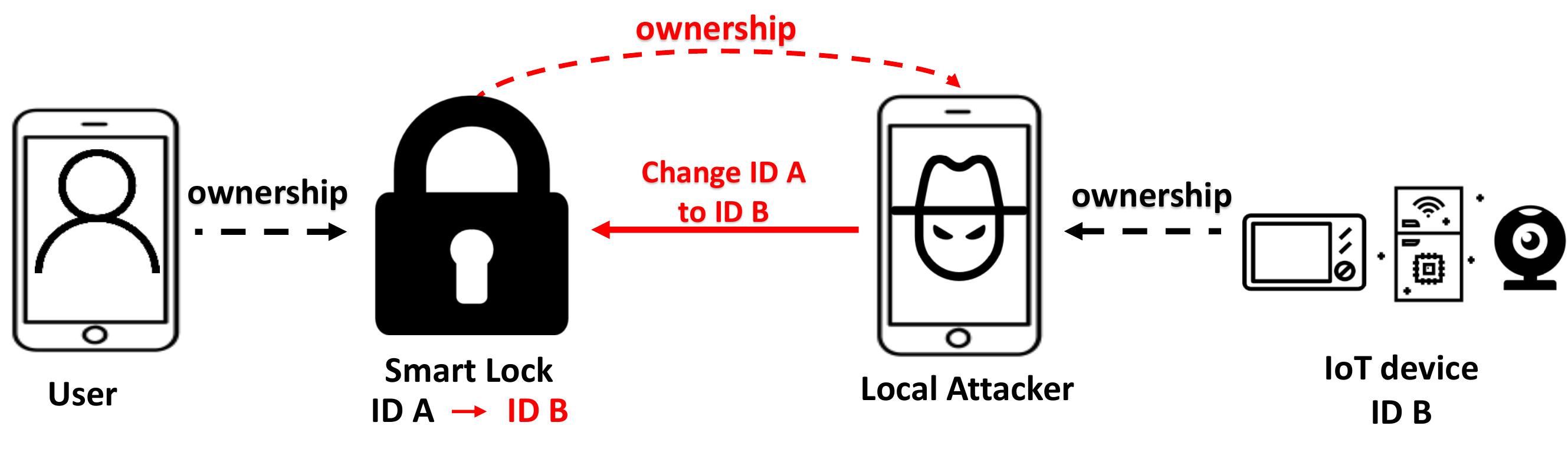}
\caption{Attack Scenario of Bug 5} 
\label{fig:bug12}
\end{figure}

\paragraph{B. Attack Scenario.}
As shown in Figure~\ref{fig:bug12}, a legitimate user is the ownership of a smart lock with the \id $\mathcal{A}$, and an attacker owns another IoT device with the \id $\mathcal{B}$.
At this point, 
if the attacker has access to the same local network with the user's device, 
he is able to send a \emph{set\_device\_id} command to the smart lock,
changing the \id of the smart lock from $\mathcal{A}$ to $\mathcal{B}$ which has been bound with the attacker's account as revealed in recent research~\cite{yao2019identifying}.

Since the \id is used to
uniquely identify the device 
and the IoT platform uses it to maintain the binding relationship
between device and owner. 
Once the \id of the device has changed,
the ownership of this device will be shifted with it.
Thus, in the above attack scenario,
after the \id of the smart lock had been changed as $\mathcal{B}$,
the attacker can illegally occupy this device forever.

\paragraph{C. Cause Analysis.}
In the above attack example,
the command \emph{set\_device\_id} which should only be sent by the remote IoT cloud has been accidentally carried out by a local attacker.
The root cause of that is 
the developers 
use common functions to deal with command sets
belong to various interactive entities in real-world IoT firmware.

Due to these common functions,
IoT firmware images often contain various execution paths from caller functions of different interactive entities
but finally to the same task function.
As shown in Figure~\ref{fig:bug12m},
if IoT firmware uses the same function to extract commands from mobile app and cloud,
except for the normal execution path from caller function 2 to task function C,
there also exists an unexpected execution path from caller function 1 to task function C.
That violates the privilege separation rules.
Thus, the local attackers are able to perform some remote sensitive command C (e.g., setting \id or unbinding the devices) which should only be sent by cloud.
Such unexpected execution paths are called \ps vulnerabilities in paper~\cite{yao2019identifying}.

\paragraph{D. Identifying Method.}
Based on the root cause of the attack,
the key to identify \ps vulnerabilities is to
identify the over-privileged common functions which will be used for
performing one command but could be invoked by different interactive entities. 
Yao et al.~\cite{yao2019identifying} developed a useful tool to identify the over-privileged common function according to the path constraints generated by symbolic execution.

\paragraph{E. Defense.}
The strict \textbf{privilege separation model} 
should be implemented in IoT firmware to make the control flow and data flow of handling commands sent by different interactive entities strictly separated.






\section{Working State Out of Synchronization}
\label{sec:bug7-8}
\begin{figure}[t]
\includegraphics[width=\columnwidth]{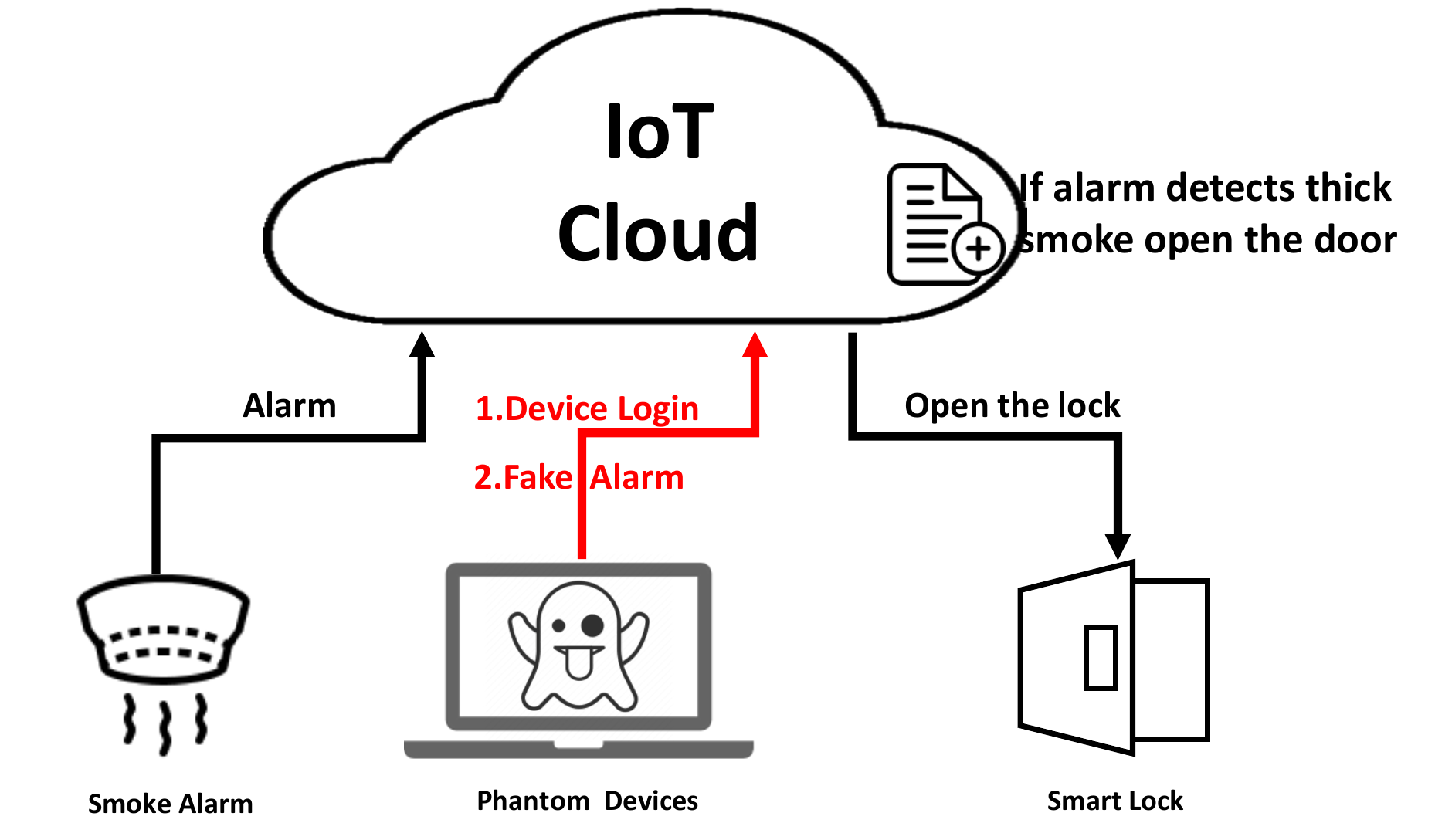}
\caption{Attack Scenario of Bug 6} 
\label{fig:bug1}
\end{figure}

\subsection{System Model}
Before the IoT devices can be securely used in a smart home,
three entities (i.e., the device, mobile app, and IoT cloud)  involved in the IoT platform need to go through several setup steps
(e.g., device discovery, device registration, device binding, etc. as mentioned in Section~\ref{sec:lifecycle}).
The three entities in different steps must stay in a legal working state 
or state combinations.
In an ideal situation,
different steps should be invoked when three entities in a different specific working state.
For instance,
the device login request should only be sent
when the cloud has already accepted device binding request but the device has not built the connection with cloud.

In addition,
the interactions between three entities will cause a
transition of their working states.
Thus,
the working states of each entity are not independent but closely related to each other.
That means an interaction may cause the working state of three entities to change altogether.
For example,
in normal operations,
if a user does not want to use his device,
he should reset and unbind the device.
IoT cloud will revoke the ownership of original and
disconnect with the device.
The working state of three entities will go back to their initial state at the same time.
After that, if anyone wants to re-use the device,
the three entities
need to go through a complete setup process, including local device discovery, device binding, and device login.

\subsection{Attack Scenario}
\paragraph{Bug 7: Insufficient State Guard.}
\label{sec:bug7}
As shown in Figure~\ref{fig:bug1},
an automation app has a home automation rule that connects a fire alarm and a smart lock, so that
in case of a fire, the alarm can detect thick smoke and
send a command to
the smart lock to open the door. 
However, Zhou et al.~\cite{zhou2019discovering} found
the attacker is able to log in a phantom device the has the same device identity as the smoke alarm.
Then the attacker can 
send fake smoke alarms via phantom device
to the IoT cloud.
As a result, the cloud will also unlock the door allowing the attacker to enter the room.

\begin{figure}[t]
\includegraphics[width=0.9\columnwidth]{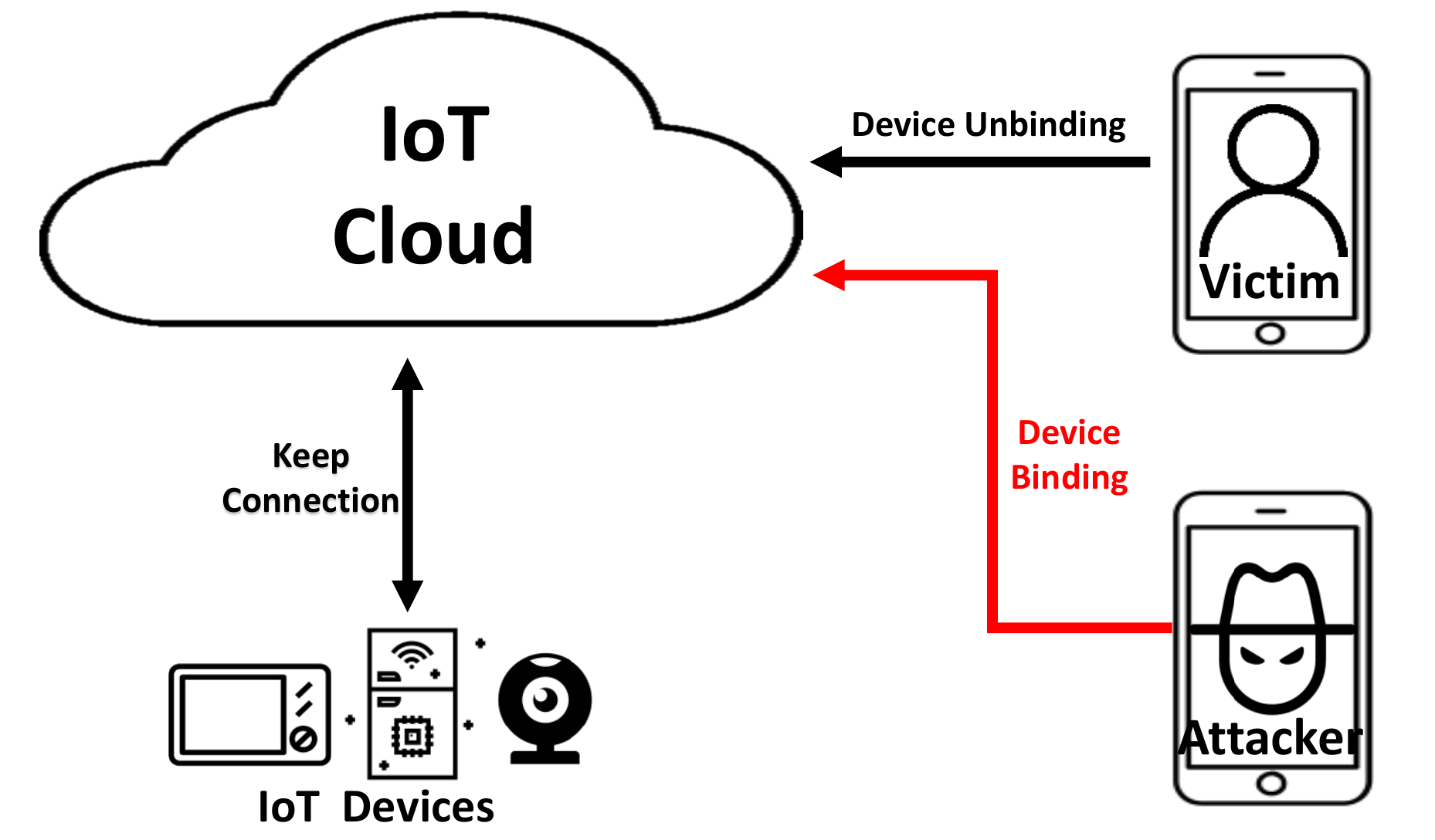}
\caption{Attack Scenario of Bug 7} 
\label{fig:bug2}
\end{figure}

\paragraph{Bug 8: Illegal States Combination.}
\label{sec:bug8}
If a user only unbinds the device but forgets to reset the device,
the IoT cloud will also revoke the ownership with the user,
but the device is still in its original state and keeps a connection with the IoT cloud.
This allows a remote attacker to forge and send a binding request with his account to cloud at that moment as shown in Figure~\ref{fig:bug2}.
Since the connection between cloud and device is still maintained,
after the cloud accepts the binding request, 
the victim's device will be directly under the control of the remote attacker without finishing other setup steps
including device discovery or logging in the cloud. 

\subsection{Cause Analysis}
In the bug 7 exploitation example, 
the IoT cloud not only just does not carry out strict device authentication, 
more importantly,
IoT cloud accepts device requests without checking its working state.
Thus, even if IoT cloud fails to distinguish the real device
with the phantom device,
as long as it is aware that the same device is still keeping the connection,
it should still refuse the same device login request based on its current working state.
Unfortunately,
many popular IoT platforms do not maintain the working state of
interactive entities.
That means most requests can be invoked at any time, which
will lead the inconsistency in the IoT cloud.
This inconsistency could further be taken advantage of, causing serious security and privacy violations.

In the bug 8 exploitation scenario,
when the cloud accepts the unbinding request, it, together with the mobile app will go back to the initial state
but the working state of the device has not been changed.
Thus, the cloud will allow the device binding request and directly transfer to the normal working state,
even if the real device has not ever been set.

\subsection{Identifying Method}
The legitimate interaction between the three kinds of entities can be
clearly represented by a working state machine and the legitimate 3-tuple state combinations of three entities can also be clearly identified according to normal operations.
Then, according to the transition rules,
unexpected interaction requests which should not happen in its current working state and unexpected state combinations can be easily identified.

\subsection{Defense}
To prevent the three entities from accepting unexpected interaction requests,
each entity should add the working state field in each communication message.
As such, the sender and receiver entity can verify if its current state allows the request to be sent out or accepted.

On the other hand,
in case three entities stay out of the legitimate state combinations, 
the IoT cloud of a platform should be responsible for synchronizing the three entities to ensure that three entities always remain in a legitimate state combination.
Since intermittent network conditions may
make it difficult to keep the three entities' working state synchronized at all time,
as an alternative solution,
the handshake protocol can be used for state synchronization.
If an unexpected state combination occurs,
the three entities should roll back to their previous legitimate state combination immediately.

\section{Sensor Data Out of Synchronization}
\label{sec:bug9-10}
\subsection{System Model}
\begin{figure}[t]
\includegraphics[width=0.95\columnwidth]{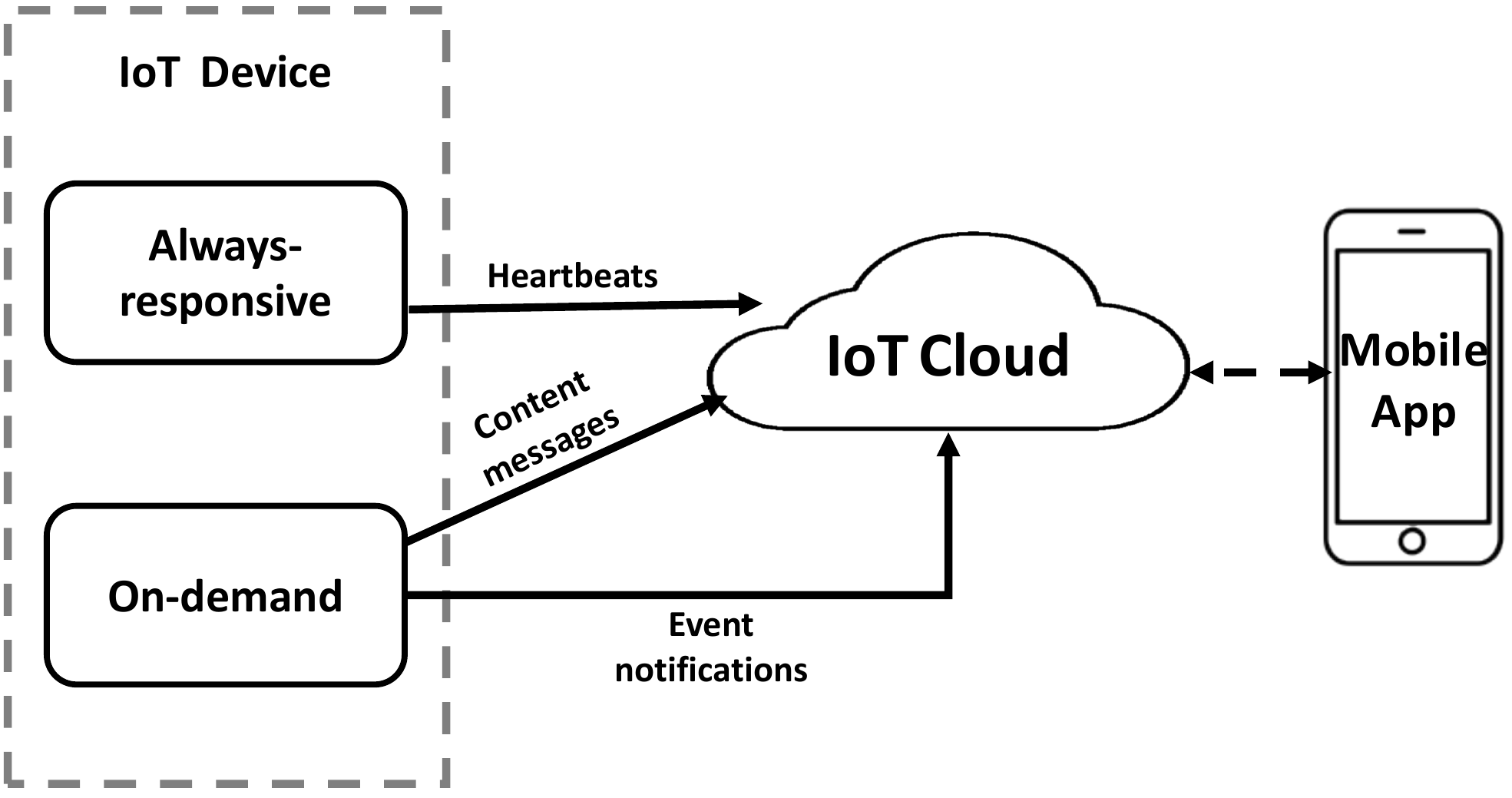}
\caption{System Model of Bug 9 and 10}
\label{fig:bug17-1}
\end{figure}

As mentioned in Section ~\ref{sec:comm},
corresponding to three kinds of device uploading message,
communication between the IoT cloud and device can be divided into three sub-channels.
It is a common practice
the three sub-channels are separated on
that the protocols, transport layer ports, servers, or time-shared on the same network flow.
In addition,
the heart-beat message and event notifications are low-bandwidth messages, and the content messages are high-bandwidth messages.
As shown in Figure~\ref{fig:bug17-1},
heart-beat messages are used for checking the availability of the device, so that this sub-channel is always-responsive.
Correspondingly,
content messages used for command responses and event notifications
are on-demand.
Note that sensor data mentioned in this system model not only represents
the sensory measurements which are the data collected by sensors, including motion detection, video recording, auditory, water detection, or other environmental sensors,
but also includes
the actuator states which are the physical status of devices (e.g. smart-lock with the locked or unlocked state).

\begin{figure}[t]
\includegraphics[width=0.95\columnwidth]{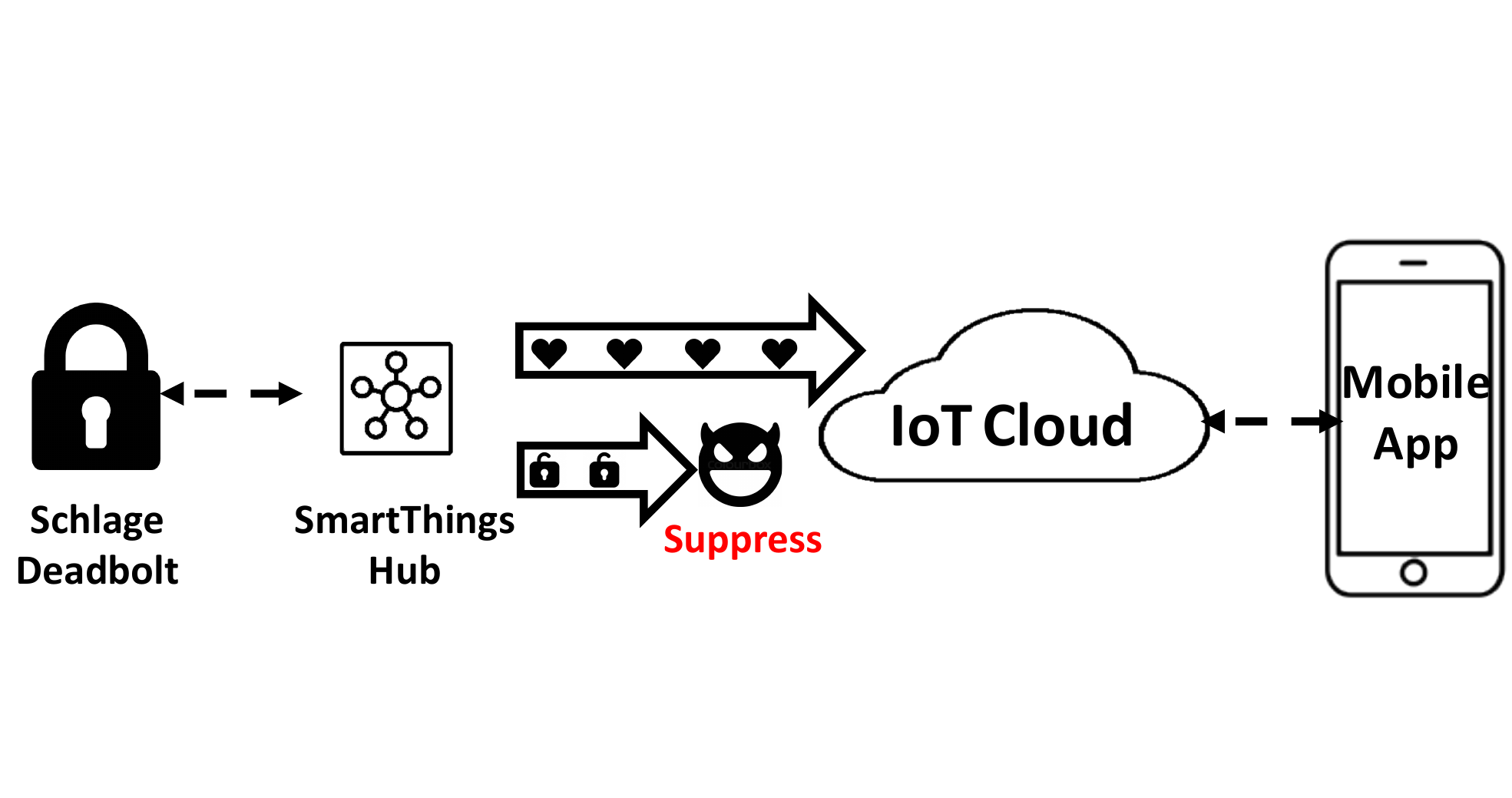}
\caption{Attack Scenario of Bug 10}
\label{fig:bug17-2}
\end{figure}

\subsection{Attack Scenario}
An attacker can exploit the out-sync of sensor data to cause sensor blinding or state confusion~\cite{oconnor2019blinded}.
Sensor blinding attack could destroy the availability of sensor devices by 
preventing the delivery of sensory measurements to the IoT cloud.
State confusion attacks the integrity of actuator state of devices reported to IoT cloud, 
and causes the state displayed in the companion mobile app to be inconsistent with the actual state of the actuator.
We assume that the attackers can fingerprint the IoT devices being used and learn the telemetry channel model for a specific device,
they can also selectively suppress a particular channel in the following attack scenarios.
Attackers can achieve this by physical layer suppression (e.g. jamming) and local network layer suppression (e.g. controlling over the wireless router).

\paragraph{Bug 9: Sensor Blinding.}
\label{sec:bug9}
As a concrete example mentioned in paper~\cite{oconnor2019blinded}, the Merkury Security Camera is a connected surveillance camera that is able to record abnormal motions such as home invasion, and uploads motion notification to AWS servers using a plain-text MQTT connection. Meanwhile, the device separately sends heartbeats (connectivity health and video content) to the AWS servers over SSL. 
Attackers can easily identify the always-responsive and on-demand messages by correlating the packet timings and blind the sensor from delivering abnormal messages. 
The server regards the device as online because it receives the periodic heartbeats, but it will not be aware the device fails to report on-demand messages. Thus, the companion mobile app will not alert the user for the abnormal event even though the device upload videos over the SSL connection.
Because the device does not buffer undelivered events if it is not disconnected, users will not get a notification even after connection reconstruction. Attackers can utilize this logic bug to eliminate forensic evidence to gain physical access to areas.

\paragraph{Bug 10: State Confusion.}
\label{sec:bug10}
In another attack scenario~\cite{oconnor2019blinded},
the author discovered state confusion in the Schlage Deadbolt.
The Schlage Deadbolt offers Z-Wave connectivity and supports different hubs including SmartThings, Iris, Alexa, etc.
After the deadbolt and the hub are paired, the user can remotely control and monitor the state of the deadbolt through the mobile app.
The author found that the on-demand channel packets were quite larger than the always-responsive heartbeat messages.
Figure~\ref{fig:bug17-2} describes an attack scenario when the deadbolt works with SmartThings hub.
An attacker could drop packets larger than 359 bytes for the SmartThings hub by local network layer suppression to prevent the transmission of unlocked state change when the deadbolt triggers an unlocked event.
After stopping the suppression, the companion mobile app reported the deadbolt as locked.
However, after a maximum period (e.g., 100 seconds) of heart-beat messages, the companion mobile app updated the state as unlocked.
Another, when the deadbolt works with the Iris hub, the attacker could suppress the on-demand channel and drop packets larger than 250 bytes.
In this scenario, the Iris hub's companion mobile app has always falsely reported the deadbolt as locked.
So for SmartThings hub, an attacker can use this short 100 seconds state confusion to sneak into the house.
And for Iris hub, the serious thing is that the attack would permanently confuse the mobile app.

\subsection{Cause Analysis}
The sensor data out of synchronization is due to packet loss and delay in the telemetry channel.
There are two main reasons for packet loss and delay.
First, the IoT device's limited storage and battery constraints feature cause a lack of on-demand event buffering and lengthy timeout periods.
Second, when an adversary suppresses on-demand messages and causes disconnection,
the IoT device typically are designed to re-establish the connection by always-responsive sub-channel and discards the event instead of re-sending the buffered event notification.

Furthermore, there have another special reason for state confusion.
IoT platform treats a state change as a single fixed event and only also devices to report the state change when the action physically occurred.

\subsection{Defenses}
There are three main methods to defend against telemetry channel suppression attacks in IoT~\cite{oconnor2019blinded}.
The first method to defend against the attack is to obscure messages sent from the devices by manipulating traffic. Another method is to establish pre-IoT virtual private networks to prevent attackers from inferring traffic activities and selectively suppressing the on-demand sub-channels. Besides, unifying the on-demand and always responsive sub-channels, redesigning priority buffer scheme, and reducing timeout length to achieve a secure IoT design are great solutions for this logic bug.





\section{Unexpected trigger action IN \sapp}
\label{sec:bug11-14}

\subsection{System Model}
\begin{figure}[t]
\includegraphics[width=\columnwidth]{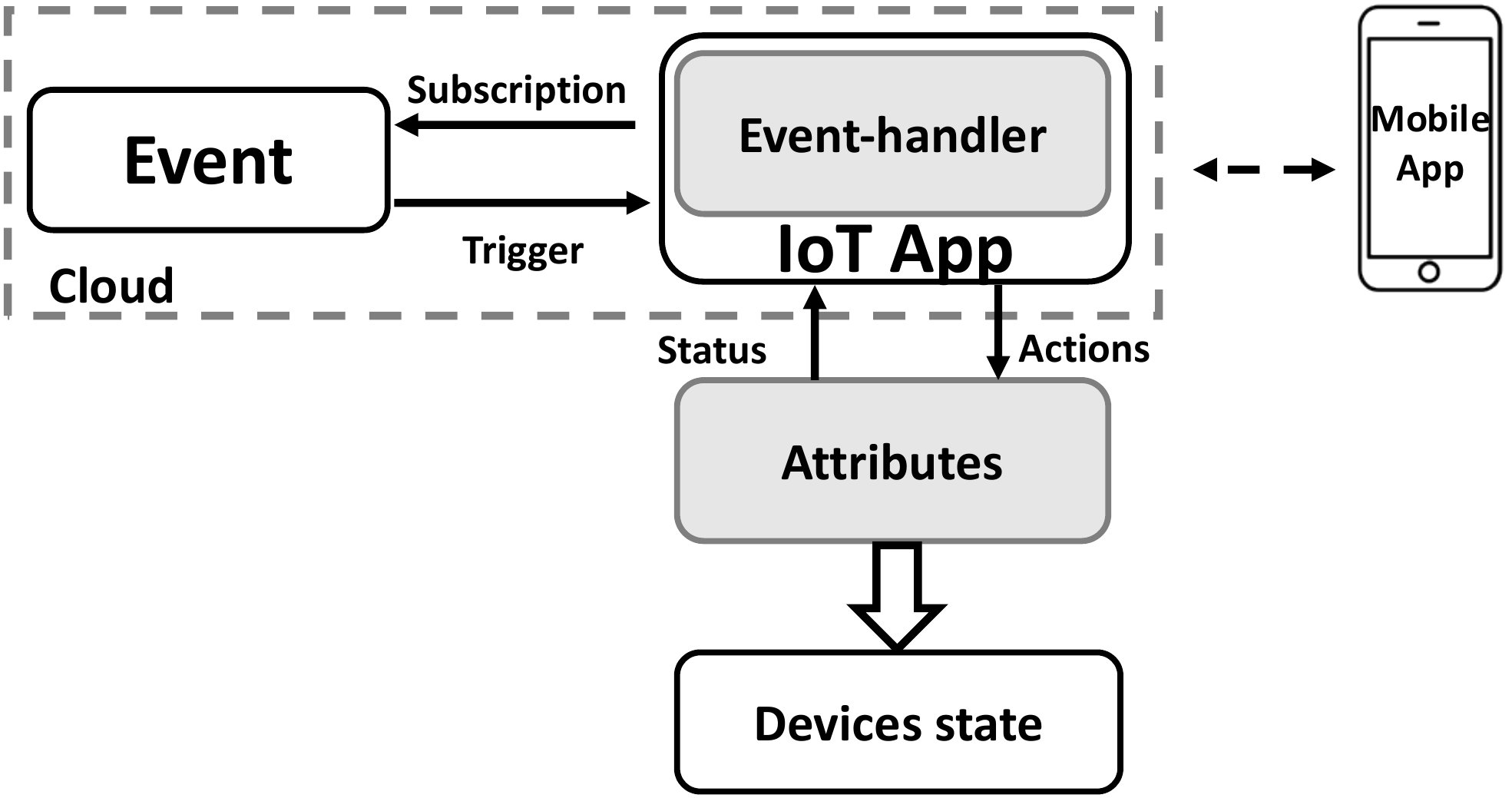}
\caption{System Model of Bug 11-14} 
\label{fig:section7}
\end{figure}

The automation app developments are based on a software stack provided by IoT platforms to realize monitoring and controlling on IoT devices \footnote{All the unexpected trigger-action bugs covered in this section lie in Samsung's SmartThings Platform}. Under the hood, as shown in Figure ~\ref{fig:section7}, the trigger-action model of the IoT platform consists of events, event-handler methods of automation app, actions, and the attributes which represent the state information of devices.
To realize the trigger-action services,
the automation app needs to register an event-handler with a device event or pre-defined event. 
The handlers are triggered to take action when these events occur. Actions represent the commands to control device states, which cause modifications on attributes, e.g., device state changes.

By exploiting the logic flows of the trigger-action rules, several critical bugs have been
disclosed by Celik et al.~\cite{celik2018soteria},
In what follows, we introduce four representative bugs in this category.

\subsection{Bug 11: Race Conditions of Events}
\label{Race}

\label{sec:bug11}
\paragraph{A. Attack Scenario.}
As is defined in~\cite{celik2018soteria}, an attribute of a device cannot be modified to conflicting values by two or more non-complementary event handlers of multiple apps working in concert, which may lead to a potential race condition. For example, ``When motion is detected, turn on the switch'' and ``Every day at midnight, turn off the switch'' will conflict if motion is detected at 12 pm. It is notable that the authors~\cite{celik2018soteria} do not investigate what attacks the adversary may realize by utilizing these bugs. Thus, the consequences caused by this bug are limited to leading devices trapped into insecure or unsafe states.

\paragraph{B. Cause Analysis.}
Upon its subscribed events' occurrence which is different from each other, two or more independent event-handlers of multiple apps are invoked to take actions possibly at the same time to manipulate the same attribute of one device to conflicting values. The sequence and timing of actions of these event-handlers usually make the final states of devices unpredictable.

\paragraph{C. Identifying Method.}

The authors who discovered the above logic bug also proposed the identifying method named SOTERIA~\cite{celik2018soteria}. This approach translates the source code of an automation app into an intermediate representation (IR) at its initial phase. With the IR being fed into the second phase, a state model of the app including its states and transitions is constructed. For the third phase, a series of IoT properties are identified for further security analysis. Property S.4 states this race condition of events bug~\cite{celik2018soteria}. Model checking is performed to find the existence of property violations when the app functioning independently or working collectively with other apps. S.4 is violated during the interaction between multiple apps by invoking actions with different device events which manipulate the same attribute of the dedicated device to conflicting values~\cite{celik2018soteria}.

\paragraph{D. Defense.}
IoTGuard~\cite{celik2019iotguard} is a follow-up work that enforces a policy checker on a dedicated server. It is comprised of three components: a code instrumentor, a data collector, and security service.
The code instrumentor provides two functions by adding extra logic to an app’s source code. One is to collect runtime information including the app’s actions, the event to trigger the action, the condition to be satisfied for the action, and the involved numerical-valued attributes, followed by sending the collected action’s information to the data collector. The other one is to insert a guard, essentially waiting for a decision from the security service on the action to be taken. 

The data collector receives all the actions' information from the instrumented app when its event-handler gets invoked, which are loaded into the dynamic model. 
The design of this dynamic model is to emulate the logic of either an app execution not interacting with other apps or unified interaction of multiple apps before an action for further security service evaluation. 

The security service is based on identified policies which are extensions of IoT properties~\cite{celik2018soteria}. The policy enforcement is actually enforcing the dynamic model to conform to the established policies. For the race condition of events, users are required to choose which action to be blocked since the nature of this bug is conflicting values of the same attribute. The decision is fed into the above guard to continue app execution, which successfully prevents the device from being stuck into unsafe and undesired states~\cite{celik2019iotguard}.

\subsection{Bug 12: Attributes of Conflicting Values}
\label{sec:bug12}

\paragraph{A. Attack Scenario.}
In a scenario where multiple automation apps are used in combination, multiple apps may change an attribute of the device to conflicting values based on the same event.
For example, App1 sounds the smoke alarm and turns on the light when the smoke is detected, and
App2 turns off a light switch when the smoke is detected.
In this case, there will be unpredictable results, that is, the light may be on or maybe off.

\paragraph{B. Cause Analysis.}
There are two major causes of this logic bug.
First, multiple apps share the same device attributes, and different apps use the same event.
Second, app designers are limited to considering a single app's program logic, and it is hard to think of the global logic in a mixed-use scenario of multiple apps.

\paragraph{C. Identifying Method \& Defense.}
Identifying methods and defense methods are similar to Section \ref{Race}.
For multi-apps, SOTERIA builds a union state model.
SOTERIA uses a safety property to identify the attributes of conflicting values.
The property states that a handler must not change attributes to conflicting values on the same event in multi-apps.
If an app does not conform to this property when running interacting with other apps, the multi-apps have attributes of conflicting values bug.
Similarly, IoTGuard uses security service to defend against this logic bug.
The security service is the same as the safety property mentioned above.

\subsection{Bug 13: Attributes Duplication}
\label{sec:bug13}
\paragraph{A. Attack Scenario.} 
The duplication of an attribute can be invoked by the same or different event handlers. When one IoT device receives the duplicate attributes, it may cause unexpected results. As an  example mentioned in~\cite{wang2019charting}, App1 calls users when their calendar receives an appointment, while App2 adds a new appointment in the user's calendar if they missed a call. Hence, if one call is missed, there will be pointless appointments filled in the user's calendar. If the triggers of one attribute are two complement event handlers, it will be a special case of an inconsistent event. For example, App1 opens a device when the motion event handler is active, while App2 is designed to open the device when the same handler is inactive. It happens when multiple applications control one device. Since there is no agreement on the logic design between different applications, they might utilize a device in the same way with divergent event handlers. 

\paragraph{B. Cause Analysis.} 
Two circumstances can trigger attributes duplication. First, this logic bug happens when multiple apps interact with the same device. Developers may publish applications controlling the same devices with different goals. Thus, an event handler updates cyclically one device with the same attributes. Second, some apps use general event handlers instead of a specific sub-event handler, such as Turn off all devices vs. Turn off the device or Disarm all cameras vs. Disarm a camera. So the general event handlers will update all corresponding attributes. Thus, one device controlled by the general event handlers and its sub-event handlers may receive duplicated attributes. 

\paragraph{C. Identifying Method \& Defense.}
Identifying methods and defense methods are similar to Section \ref{Race}. The identifying method of this logic bug is to create a union state-model of interacting apps. By extracting the complete behaviors when running the multiple apps, we can identify the attributes duplication. An attribute cannot be changed repeatedly to the same value by the same or different event handlers. If one app violates this property, it has the attributes duplication bug. To defend against this logic bug, when multiple apps implement the same functionality by changing one attribute to the same values, the data collector adds parallel edges from the event state to the action state and labels the edges with
the app’s objects. In this case, a policy is defined by security to prevent repeated action.

\subsection{Bug 14: Missing Events}
\label{sec:bug14}
\paragraph{A. Attack Scenario.} 
Under the trigger-action model of the automation app programming framework, an event must be subscribed by the event handler
whose code contains logic that handles that event.
However, if (1) a handler takes an
event-type value but performs different actions
according to the types of events, or (2) the handler has a case for handling event, but the
app does not declare that the handler subscribes
to the event, the expected action cannot be taken.
For example, a smart lock app is supposed to
unlock the door when the user approaches it and lock
it when the user is away. This requires the app
to subscribe to the location mode change event handler.
If this is not done, the app would fail to function.

\paragraph{B. Cause Analysis.} 
This bug roots in the fact that the developers often do
not strictly follow the programming paradigm in the smart home development.

\paragraph{C. Identifying Method.} 
The bug is disclosed by using the same methodology as described in
Section~\ref{Race}.
Concretely, from the extracted state model,
if an event is found to have
zero subscribers, it is likely to be a missing
events vulnerability.

\paragraph{D. Defense.} 
To avoid this from happening, on the one hand,
developers should be trained to understand
the basic model of smart home programming and follow the best practice.
On the other hand,
currently, defenses including IoTGuard are very ineffective
because no general policies can be programmed as part of the security service --
no defense can be enforced when no action
is requested at all.






\section{Information Flow Hijacking in \sapp}



\label{sec:bug15-16}

\subsection{System Model}
To provide more home automation apps for user,
most IoT platforms allow the user install the   
automation applications from the third-party trigger-action services and one of most
popular is IFTTT.
The automation apps in IFTTT service app market named applets.
One applets is consist of three major components (trigger service, action serivce and filter code).
The triggers and actions of IFTTT applets
can be provided by different partners' services.
Between the trigger and the action, there is filter code, which is JavaScript code snippets.
It can use the APIs provided by the partners' services and customize the output of the applets.
For example, the filter code can obtain and customize the URLs generated by the IFTTT cloud service for the uploaded data from the trigger service.
Then the URLs will be passed to the action service.

\subsection{Attack Scenario}

\paragraph{Bug 15: URL-Based JS Injection.}
A malicious applet advertises itself as a photo backup tool, which back ups iOS photos to Google Drive.
As long as the photo is taken, it is uploaded to the IFTTT cloud and a URL is generated for this photo, which would be accessed by the action code of Google Drive to retrieve the photo.
However, in the filter code of this applet, there is a malicious JavaScript code snippet, which manipulates the URL of the uploaded photo.
The manipulated URL links to the attacker-controlled server and includes the original URL as a parameter part.
When the URL is accessed by the action code from Google Drive, the attacker-controlled server will be accessed.
In the end, it can get the photo through the URL in the parameter part, and send the photo to Google Drive without the user's attention.

\paragraph{Bug 16: URL-Based HTML Tag Injection}
A malicious applet advertises itself as a notes-to-email tool, which records a list of notes to the user's email.
The notes can be taken through Google Assistant, etc.
When a note is taken, the filter code of the malicious applet injects an invisible HTML markup tag, with a URL linking to the attacker-controlled server, into the note.
The URL includes the note content.
As soon as the email is read by the user, the attacker-controlled server would be accessed with the note content as a part of the URL, which results in the privacy leakage.

\subsection{Cause Analysis}
The root cause of this problem is IFTTT service does not prevent the information flow from the private source to public sinks.
That is, the information from the private source should not go to arbitrary public places.
For example, in the above example, the photo should not go out of iOS system, IFTTT cloud and Google Drive.
However, without the restriction, the malicious applet can send the URL of the photo generated by the IFTTT service to other servers, which results in the leakage of the private information.

\subsection{Identifying Method \& Defense }
Bastys et al. propose two solutions to defend the attacks caused by the URL-based information flow in IFTTT service, breaking the flow and tracking the flow~\cite{bastys2018if}.
Breaking the flow means to classify the trigger and action service providers, and restrict the sources and sinks to either exclusively private or exclusively public data.
In this way, there is no flow from private to public, thus preventing privacy leakage.
Specifically, the access to public URLs in the filter code is disabled or delegating the choice to the users at the time of the applet's installation.
However, both methods for breaking the flow may over-kill the benign applets and is not flexible for future service features.
On the other hand, tracking the flow ensures the only way to include links or markup on the action-based APIs is through using API constructors provided by IFTTT service.
By monitoring the information flow in the applet, this method can prevent privacy leakage and eliminate the defects in breaking the flow method.







\section{Vulnerable Task Management in RTOS}
\label{sec:bug17}
\subsection{Bug 17: Lack of Isolation between Context Table and Tasks}
\begin{figure}[t]
	\includegraphics[width=\columnwidth]{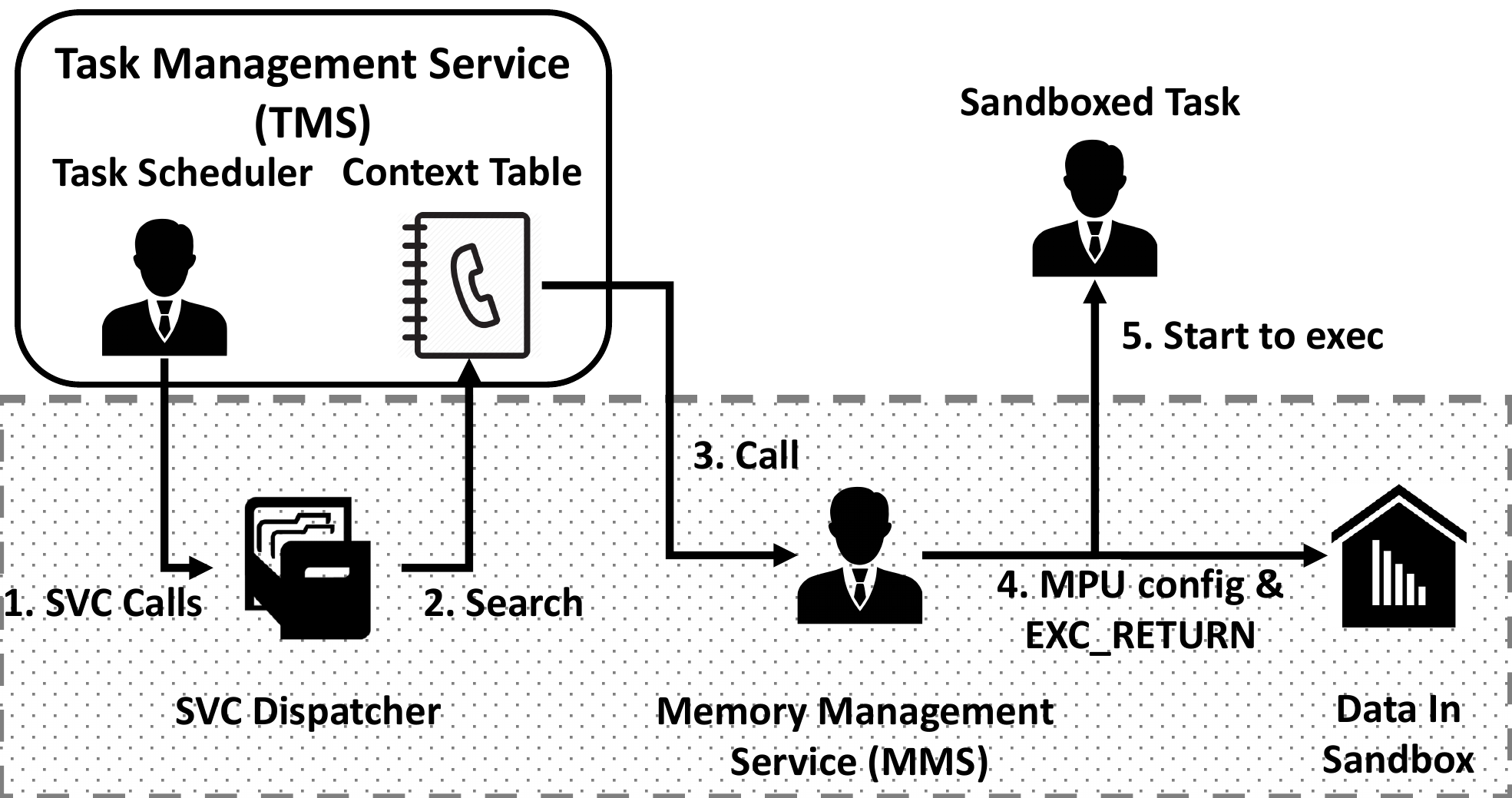}
	\caption{System Model of Bug 17}
	\label{Fig.sub.1}
\end{figure}
\begin{figure}[t]
	\includegraphics[width=\columnwidth]{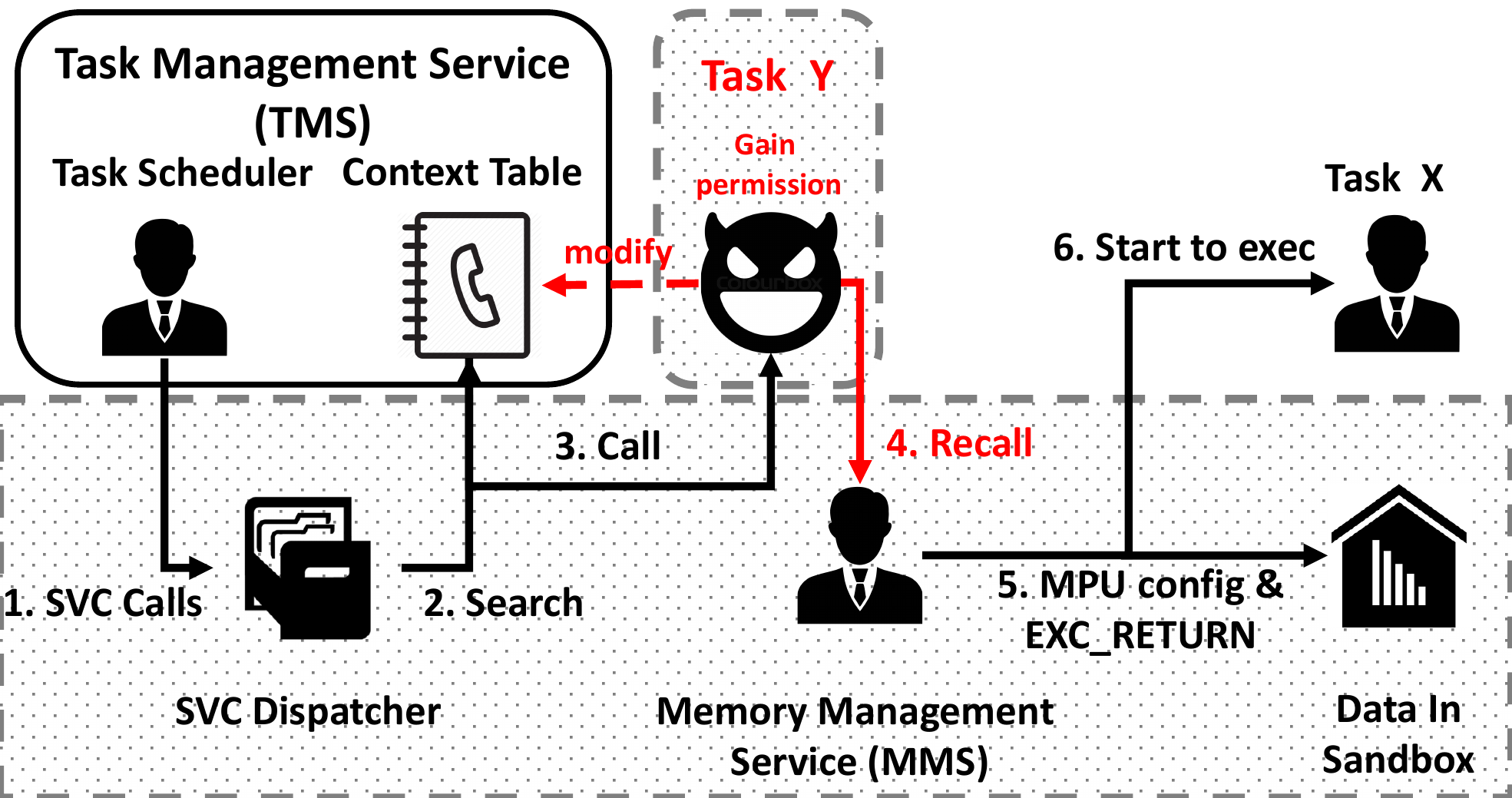}
	\caption{Attack Scenario of Bug 17}
	\label{Fig.sub.2}
\end{figure}
\paragraph{A. System Model.}
Arm Mbed, a representative commercial RTOS for IoT devices, designs uVisor~\cite{MbeduVisor} for task sandboxing.
The Memory Management Service (MMS) offered by uVisor always runs in the privileged mode while the Task Management Service (TMS) offered by Mbed and all tasks (including both sandboxed and unsandboxed tasks) run in the unprivileged mode. 
The Context Table, which holds a set of address pointers pointing to particular memory management functions in MMS, is used by the TMS to index the ``services'' provided by the MMS. 
The Mbed task sandboxing mechanism ensures that the data of every sandboxed task will be stored in memory and only accessible in the privileged mode, and that the memory access permission can only be switched by the MMS. As an example of TMS, the task scheduler is finding the ready-to-start task shown in Figure~\ref{Fig.sub.1}. 
Were it to be a sandboxed task, the task scheduler uses pre-defined SVC calls to trigger SVC dispatcher to run in the privileged mode. The dispatcher then searches the context table to call MMS, which configures the MPU to set the memory region of the task's data with readable/writable permissions in the unprivileged mode. Finally, the dispatcher uses specific instructions (e.g., load EXC\_RETURN into the PC register in Cortex-M processors) to return to the unprivileged mode and gives control to the sandboxed task.

\paragraph{B. Attack Scenario.}
There are two tasks shown in Figure~\ref{Fig.sub.2},
the sandboxed task X with its secure data and the unsandboxed task Y. 
When task X runs, an attacker controls a specific task (i.e., task Y) to manipulate the context table to replace one of its pointers with one of task Y's functions to access the secure data.
Then every time the dispatcher searches the context table to call a memory management service during task scheduling, it actually calls the (malicious) function in task Y. Because the dispatcher hasn't used instruction EXC\_RETURN yet, task Y will run in the privileged mode and gain the permission to access the secure data. 
After that task Y can continue to call the intended memory management service to make sure that the malicious behavior is non-perceived.

\paragraph{C. Cause Analysis.}
The IoT devices vulnerable to attacks exploiting Bug 16 only support two privileged modes (e.g., privileged and unprivileged modes). 
The TMS and tasks both run at the same privilege level and there's no mechanism to restrict tasks' accessibility to TMS, so the system actually cannot prevent its tasks from directly modifying the context table in TMS, which means that only when the context table is isolated from tasks can the system be secured. 
This vulnerability is first reported in paper~\cite{openreview:LIPS}.

\paragraph{D. Identifying Method.}
To identify if the system is in the risk of MMS hijacking attack, 
an improved identification technique with control flow matching~\cite{DBLP:conf/ccs/2016} is proposed here. 
The system designer first registers original control flow of MMS by either static or dynamic analysis methods~\cite{DBLP:conf/pldi/2005,DBLP:journals/toplas/FerranteOW87}. Then the software codes are instrumented for OS to dynamically collect the control flow. 
When an unexpected input is captured by the system, which manipulates a task accesses sandboxed data with an unregistered control flow of MMS, it will be identified and considered as hijacking the MMS. 

\paragraph{E. Defense.}
LIPS~\cite{openreview:LIPS} provides protection domains for RTOS services and tasks under the same privileged level to achieve an intra-privilege isolation, so that context table cannot be modified by tasks. LIPS is incorporated into uVisor to not only keep its security but also achieve the dynamic switching of tasks' memory accesses.

\begin{figure}[t]
	\includegraphics[width=\columnwidth]{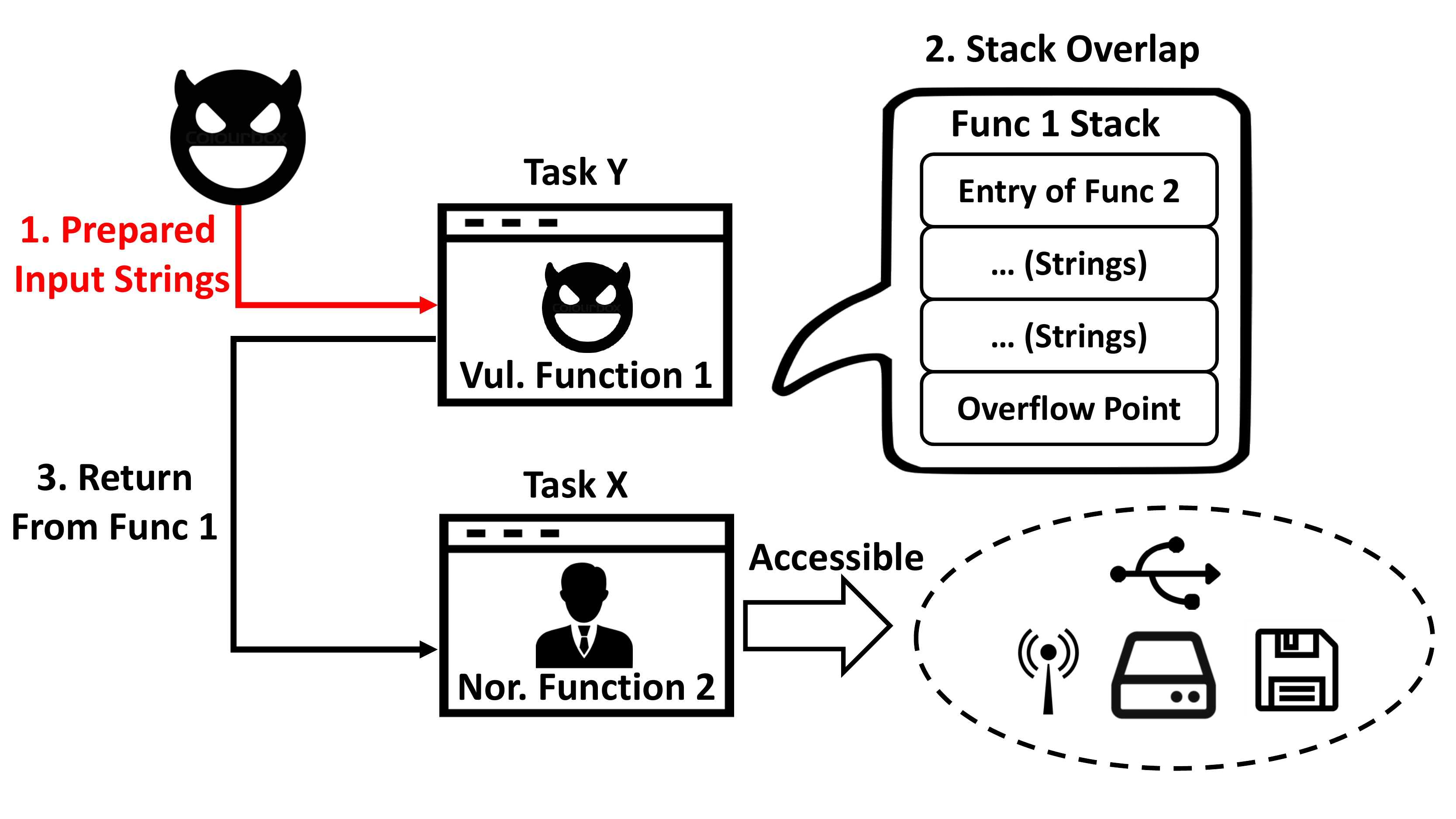}
	\caption{Attack Scenario of Weakness 1}
	\label{fig:Task-flow-redirecting}
\end{figure}

\subsection{Weakness 1: Inadequate Task Memory Isolation}
\label{sec:weak}

\paragraph{A. System Model.}
In x86 processors, every task could run in an isolated virtual memory address space offered by the virtual-to-physical address translation of MMU. 
Through mechanisms such as Inter-Process Communication (IPC) and shared memory, tasks can exchange information but still be restricted to their own address spaces. Since the MPU does not support virtual memory, the RTOSs deployed on IoT devices simply layout (the code and data of) tasks into a (shared) physical memory address space, which makes the memory address space a large attack surface for attackers who are exploiting a memory corruption vulnerability. 

\paragraph{B. Attack Scenario.}
As shown in Figure~\ref{fig:Task-flow-redirecting}, it is obvious that an attacker can compromise function 1 in task Y through a buffer overflow to redirect the control flow to function 2 in task X.

\paragraph{C. Cause Analysis.}
First, traditional OSs use MMU to abstract physical memory in the form of virtual memory for restricting tasks' access accessibility, while RTOSs deployed on IoT devices do not support this feature. 
Second, MPU roughly divides the physical memory into MPU regions of a fixed number, each assigned with different access permissions under both privileged and unprivileged modes. But this lightweight access control design cannot provide such isolation of the same level as that by MMU. 

\paragraph{D. Identifying Method.}
To identify this vulnerability, the attacker needs to determine the target task X who contains a function 2, whose entry address is known to the attacker. Then, the attacker finds or designs a intermediary task Y who contains a trampoline-function (e.g., function 1) with the potential ability to jump to the target address of the attack and cannot resist the memory corruption attack (e.g., buffer overflow). 
Finally, if the attacker can compromise the function 1 to call function 2, it is considered that there actually is an inadequate task isolation vulnerability.

\paragraph{E. Defences.}
Kim et al.~\cite{kim2018securing} designed a security architecture that virtually partitions the memory space and enforces memory access control of a RTOS. Through off-line analysis on identifying the reachable memory regions of a task, 
they used MPU to conduct run-time memory access control for each task and finally reduces the memory spaces which are open to attackers.

\section{Discussion}
In this section, we comment on a few things we learned from the elaborated IoT logic bugs. 

\subsection{Some Vulnerabilities are Inherited from Traditional Computing systems}

Obviously, some logic bugs root from the 
design issues of traditional computing systems, such as 
authentication bypassing and task isolation vulnerabilities.
Since in traditional computing systems there have been many
advanced defense techniques, 
we are curious of what holds back 
IoT platforms and devices from applying
off-the-shelf defenses and solutions.
We try to answer this question from 
two aspects.

\paragraph{Human-related Reasons.}
First, most IoT devices are low-cost energy-efficient devices.
However, many additional security features rely on hardware components such as physical unclonable functions (PUFs) and cryptography chips,
not only the price of SoC could be raised accordingly,
but also increased power consumption and reduced use lifetime.
Second, as IoT business continues to grow, manufacturers are
facing increased time-to-market pressure.
Although security is a concern, manufacturers tend to
reuse existing code base, which is obsolete and less tested
on the Internet.
Besides the IoT program and applications are becoming more and more complicated,
few companies are willing to harm the profit by rewriting the whole application applying security features or carrying out security tests on existing code.

\paragraph{Technical Challenges.}
Many IoT devices are powered by lightweight microcontrollers (e.g., ARM cortex-m and MSP430)
and have less memory resource,
thus instead of well-armed systems,
they can only run lightweight RTOS or even bare-metal systems.
However, most of these lightweight systems lack mitigation protections like W\^{}X and ASLR.
For instance, any code based on FreeRTOS is running in supervisor mode.
Even some microcontrollers support hardware security features like MPU~\cite{zhou2019good},
but we found few IoT platforms adapt it well. 
Furthermore, additional software-based TEE implementations introduce performance overhead.
For example, as shown in previous research~\cite{kim2018securing},
they observed that the kernel-memory-enable RTOS systems failed to meet the deadline constraints of the real-time tasks.

Overall, most IoT platforms and systems still suffer from classical logic bugs just as previous computer software does. 


\subsection{New Challenges in Securing IoT Platforms}
There are also new security challenges brought by
the unique design features of the emerging IoT applications.

\paragraph{More Entities Involved.}
Compared with traditional computing systems,
there are more entities involved.
In traditional computing systems, only the client and the server 
need to mutually authenticate each other and
the authorization is then performed based on the identification information and  access control policies.
In the IoT platform, more entities (the IoT device, the cloud and the mobile app) are involved.
This makes authentication and authorization more complicated.
If authentication or authorization are not implemented properly,
security problem may arise.
For example, as shown in Section~\ref{sec:bug7}, to send a falsified request,
the attackers only need to bypass device authentication.
Similar problems also result in Bug 3 as mentioned in Section~\ref{sec:bug4}. 



As the interactions among the three entities become complicated,
the working state management becomes necessary and should be paid more attention.
As revealed in Section~\ref{sec:bug7},
exploiting unexpected state transitions can cause serious consequences.

\paragraph{Interaction with Physical World.}
Through IoT automation services,
IoT devices interact with their surrounding environment.
This new interaction model
brings about potential security hazards.
For example,
as discussed in Section~\ref{sec:bug11-14} and Section~\ref{sec:bug9},
if users deploy several automation trigger-action apps in one smart home,
multiple indirect interactions may unexpectedly influence each other due to race conditions of the same event, missing events, etc.
The attackers can take advantage of this design vulnerability to indirectly control the victim's devices. 

\paragraph{Unattended Usage.}
Unlike the PCs or mobile phones,
typically,
IoT users do not need physical access to the IoT devices but remotely monitor and control the devices via mobile apps or web services.
Therefore,
it is hard to verify the real-time status of IoT devices and be aware of the unexpected behaviors of devices in time.
This makes it easier for the attacker
to disguise their attack behaviors.
As described in bug 7,
since the victim can only remotely check his mobile app,
he cannot observe any unusual status of the
device in his smartphone (although it has been substituted by a phantom device).

\paragraph{Shared Devices.}
The IoT devices are often shared among family members, friends or even strangers from time to time.
For example, as mentioned in Section~\ref{sec:bug1},
the smart home devices have been widely used in rentals and hospitality services. 
In a previous study, it was found that 60\% of guests would actually pay more for a vacation rental home with a smart home feature~\cite{airbnbsurvey}.
This makes it easier for the attackers to get local or even physical access to the devices.
Thus,
hard-coding sensitive device information like credential and identification
information become risky for IoT devices,
even if the credential is long enough and unpredictable.

\section{conclusion}

In this paper, we provide a review of the recently 
discovered 
logic bugs that are specific to IoT platforms and systems. 
In particular, 17 logic bugs and one weakness falling into seven categories of vulnerabilities are reviewed in this survey, and the seven categories are as follows: authentication problems,  
over-privileged capabilities, 
out of synchronization at platform-level,  
sensor data out of synchronization, 
unexpected trigger actions in IoT apps, 
unexpected code injection, and 
task sandboxing vulnerabilities in RTOS.

\bibliographystyle{ACM-Reference-Format}
\bibliography{0.main}

\end{document}